\documentclass[12pt]{article}
\usepackage{epsfig,amsmath,amssymb,mathrsfs}
\usepackage[utf8]{inputenc}
\usepackage[colorlinks=true,citecolor=blue,,linktocpage=true,linkcolor=blue,urlcolor=black]{hyperref}
\usepackage{mathtools,slashed}
\usepackage{array}
\usepackage{xcolor}
\usepackage[force]{feynmp-auto}
\usepackage{caption}
\usepackage{subcaption}
\usepackage{overpic,youngtab}
\usepackage[latin,english]{babel}
\usepackage{amsmath}
\usepackage{amssymb}
\usepackage{epstopdf}
\usepackage{graphics,psfrag,rotating}
\usepackage{mathabx}
\usepackage{graphicx}
\usepackage{dcolumn}
\usepackage{float}
\usepackage{pdflscape}
\usepackage{array}
\usepackage{booktabs}
\usepackage{amscd}
\usepackage{mathtools}
\usepackage{fancybox}
\usepackage{fix-cm}
\usepackage[colorlinks=true,citecolor=blue,,linktocpage=true,linkcolor=blue,urlcolor=black]{hyperref}
\usepackage{tikz}
\usetikzlibrary{matrix}
\usetikzlibrary{positioning}
\usetikzlibrary{arrows}

  {\end{list}}%

\makeatletter
\renewcommand{\section}{\setcounter{equation}{0}\@startsection
 {section}%
 {1}%
 {0pt}%
 {-1\baselineskip}%
 {0.4\baselineskip}%
 {\bfseries\large}}%
\renewcommand{\subsection}{\@startsection
 {subsection}%
 {2}%
 {0pt}%
 {-0.75\baselineskip}%
 {0.2\baselineskip}%
 {\bfseries}}%
\renewcommand{\subsubsection}{\@startsection
 {subsubsection}%
 {3}%
 {0pt}%
 {-0.5\baselineskip}%
 {0.1\baselineskip}%
 {\sc}}%
\makeatother

\DeclareMathAlphabet{\mathpzc}{OT1}{pzc}{m}{it}

\usepackage{tikz}
\usetikzlibrary{patterns,snakes}
\usetikzlibrary{shapes.misc}
\usetikzlibrary{decorations.markings,decorations.pathmorphing}
\usetikzlibrary{calc}
\tikzstyle{spring}=[line width=0.8,black,snake=coil,segment amplitude=4.25,segment length=4.75,line cap=round]

\def\be{\begin{equation}}
\def\ee{\end{equation}}

\def\g5{\gamma_{5}}

\def\id3k{\int\!\! \dfrac{d^3\!\vec{k}}{(2\pi)^3 }}

\newcommand{\bea}{\begin{eqnarray}}
\newcommand{\eea}{\end{eqnarray}}
\newcommand{\beann}{\begin{eqnarray*}}
\newcommand{\eeann}{\end{eqnarray*}}
\newcommand{\ba}{\begin{array}}
\newcommand{\ea}{\end{array}}

 \def\g {\gamma}

\newcommand{\email}[1]{\href{mailto:#1}{\tt #1}}

\begin{document}

\rightline{\scriptsize{FT/UCM 314-2024}}
\vglue 50pt

\begin{center}

{\LARGE \bf Witten diagrams in momentum space and one graviton exchange between scalars in Weyl invariant unimodular gravity.}\\
\vskip 1.0true cm
{\Large Jesus Anero$^{\dagger}$, Carmelo P. Martin$^{\dagger\dagger}$ }
\\
\vskip .7cm
{
	$\dagger$Departamento de F\'isica Te\'orica and Instituto de F\'{\i}sica Te\'orica (IFT-UAM/CSIC),\\
	Universidad Aut\'onoma de Madrid, Cantoblanco, 28049, Madrid, Spain\\
	\vskip .1cm
	{$\dagger\dagger$Universidad Complutense de Madrid (UCM), Departamento de Física Teórica and IPARCOS, Facultad de Ciencias Físicas, 28040 Madrid, Spain}
	
	\vskip .5cm
	\begin{minipage}[l]{.9\textwidth}
		\begin{center}
			\textit{E-mail:}
			\email{$\dagger$jesusanero@gmail.es},
			\email{$\dagger\dagger$carmelop@fis.ucm.es}.
			
		\end{center}
	\end{minipage}
}
\end{center}
\thispagestyle{empty}

\begin{abstract}
We tackle head-on the computation of the $s$-channel Witten diagram in momentum space corresponding to the exchange of a graviton between minimally coupled scalars in Weyl invariant unimodular gravity. By means of a lengthy calculation,  we show first that the value of the diagram in question is the same as in General Relativity and, then, we  obtain a compact  expression for it  in terms of the Mandelstam variables.
\end{abstract}

\vfill
\clearpage

\section{Introduction}

Unimodular gravity is a theory of gravity which solves the huge-radiative-correction part of the cosmological constant problem, for the vacuum energy does not gravitate in that  theory \cite{vanderBij:1981ym, Zee:1983jg, Buchmuller:1988wx, Henneaux:1989zc}. In unimodular gravity the cosmological is not a part of the classical action of the theory, so that it shows up in the classical theory as an integration constant. At  the quantum level, the cosmological constant occurs as parameter of the background field when computing the on-shell perturbative background-field effective action \cite{Alvarez:2015sba} and as a property of boundary states when computing transition amplitudes between those states  \cite{Buchmuller:2022msj}.

There are several approaches to define unimodular gravity as a quantum field theory: see Refs. \cite{Eichhorn:2013xr, Padilla:2014yea, Alvarez:2015sba, Bufalo:2015wda,  deLeonArdon:2017qzg, DeBrito:2019gdd, Baulieu:2020obv, deBrito:2021pmw, Kugo:2022iob, Garcia-Lopez:2023ulg}. It is not known whether they yield the same quantum theory, even when the background metric is Minkowski, for they involve different sets of ghosts. This is an open problem, as it is their equivalence to General Relativity when the cosmological constant is set to zero. The reader is referred to Refs. \cite{Carballo-Rubio:2022ofy} and \cite{Alvarez:2023utn} for recent reviews.

In this paper we shall employ  the formulation of unimodular gravity put forward in \cite{Alvarez:2006uu, Alvarez:2005iy}. In this formulation one solves first the unimodularity constraint ${\rm det}\,\hat{g}_{\mu\nu}(x)=-1$ by  introducing an unconstrained tensor field $g_{\mu\nu}(x)$ such that
\begin{equation*}
\hat{g}_{\mu\nu}(x)=\dfrac{g_{\mu\nu}(x)}{|g|^{1/D}(x)},
\end{equation*}
where $\hat{g}_{\mu\nu}$ denotes the unimodular metric in $D$-dimensional space-time. Then, the path integral of the theory is defined by using the standard linear splitting  $g_{\mu\nu}(x)=\bar{g}_{\mu\nu}+\kappa h_{\mu\nu}$, along with standard quantization methods. $\bar{g}_{\mu\nu}$ is the background field and $h_{\mu\nu}$ is the graviton field, the latter is integrated over in the path integral. For a discussion of the quantum inconsistencies arising when other quantization methods are used jointly with the  linear splitting just mentioned the reader is referred to ref. \cite{Alvarez:2023kuw}.
We shall call the formulation of unimodular gravity we have just quickly discussed Weyl invariant unimodular gravity; for in addition to being invariant under transverse diffeomorphims, it is also invariant under Weyl transformations of $g_{\mu\nu}$.

The computation of boundary correlators in momentum space for Anti-de Sitter (AdS) space plays an important role in the analysis and applications of Maldacena's  gauge/gravity correspondence \cite{Maldacena:1997re} --see \cite{Ammon:2015wua,Nastase:2015wjb}, for  introductions to this subject. These boundary correlators can be expressed as a sum over the so-called Witten diagrams, which were introduced in ref. \cite{Witten:1998qj}. The study of such diagrams, and the corresponding correlators, in momentum space has been a subject of research  for more than a decade --see Refs. \cite{Raju:2011mp}--\cite{Bzowski:2023jwt} for a partial list of references. It should also be noted that the computation of boundary correlators in Anti-de Sitter and de Sitter spaces is also relevant in connection with Cosmology \cite{Maldacena:2002vr, McFadden:2009fg}.

Up to the best of our knowledge, the computation of the boundary correlators mentioned in the previous paragraph when unimodular gravity replaces General Relativity has never been taken up in the  literature. This state of affairs is not better in position space: see \cite{Anero:2022rqi}, though. The main purpose of this paper is to remedy this unwanted situation by explicitly computing the $s$-channel Witten diagram --see Figure 1, which describes the exchange of a graviton between scalars on an Euclidean AdS background. Here the gravity theory will be Weyl invariant unimodular  gravity  and the scalars will be minimally coupled to the graviton field. Whether the diagram in question has the same value as in General Relativity is a non-trivial issue for the following reasons: first, in unimodular gravity the graviton field does not couple to the vacuum energy; second, the graviton field only couples to the traceless part of the energy-momentum tensor, unlike in the General Relativity case; third, the gauge symmetries of the  theory are transverse diffeomorphisms and Weyl transformations, not the full diffeomorphism group. Of course, the first two reasons have to do with the last reason.

The layout of this paper is as follows. In Section 2, we give the action of the model we shall deal with and the background metric for Euclidean Anti-de Sitter  in unimodular Poincar\'e coordinates. The boundary-to-bulk scalar propagator and the bulk-to-bulk graviton propagator in the axial gauge for Weyl invariant unimodular gravity are worked out in Section 3. Section 4 is devoted to the computation of the $s$-channel Witten diagram in momentum space corresponding to the exchange of a graviton between minimally coupled scalars in our unimodular gravity theory. In this section, we show first by means of a lengthy computation that the value of the Witten diagram at hand is the same as in General Relativity. Then, we recompute \footnote{The value of the Witten diagram in question had been computed in \cite{Ghosh:2014kba} and \cite{Armstrong:2023phb}, a fact we were not aware of until we had obtained all the results issued in this paper.} the value of that very diagram in General Relativity to obtain a compact result in terms of the Mandelstam variables. The conclusions are stated in Section 5. In Appendices A and B, we display the value of the integrals needed to carry out the explicit computations done in Section 4. Some details of our computation of Witten diagram in Figure 1 for General Relativity can be found in Appendix B. Finally, let us say that the computations displayed in this paper would not have been feasible had we not used the symbolic manipulation systems FORM \cite{Ruijl:2017dtg} and Mathematica \cite{mathematica}.

\newpage
\section{The model and its classical action}

Our model will be that of unimodular gravity minimally coupled to a massless scalar field on an Euclidean $\text{AdS}_{4}$ background. Any interaction of the massless scalar field with any other field will be of no bearing on  the computations carried out in the sequel. Hence, the classical action governing the dynamics of the our model will be the following functional:
\begin{equation}
 S_{\text{\tiny{class}}}\,=\, \frac{4}{\kappa^2}\Big(\int_{\cal M}d^4 x\,R[\hat{g}_{\mu\nu}]\,+\,2\int_{\partial\cal M}d^{3} y\,\sqrt{\hat{g}^{(b)}}K\Big)+
 \int_{\cal M}d^4 x\,\hat{g}^{\mu\nu}\partial_{\mu}\phi(x)\partial_\nu\phi(x).
 \label{classaction}
\end{equation}

Let us briefly discuss the mathematical objects occurring in the previous equation. First, $\kappa= 32\pi G$, ${\cal M}$ stands for Euclidean $\text{AdS}_{4}$ and ${\partial\cal M}$ denotes its boundary. Euclidean $\text{AdS}_{4}$ is defined as the set of points $(w,\vec{x})$,
with $0<w<\infty$ and $\vec{x}\in \mathbb{R}^3$, where a Riemannian structure is defined by the line element
\begin{equation}
ds^2\,=\,\left(\frac{L}{3w}\right)^2 dw^2+\left(\frac{3w}{L}\right)^{2/3}\delta_{ij}dx^i dx^j,\quad\quad i,j=1,2,3.
\label{UGline}
\end{equation}
The Riemannian metric  --say, $\bar{g}_{\mu\nu}$-- with non vanishing entries
\begin{equation}
\bar{g}_{ww}=\left(\frac{L}{3w}\right)^2,\quad \bar{g}_{ij}=\left(\frac{3w}{L}\right)^{2/3}\delta_{ij},
\label{BackUGmet}
\end{equation}
which defines the line element in (\ref{UGline}), will be called the background unimodular metric.

As discussed in ref. \cite{Anero:2022rqi}, the change of variables
\begin{equation}
w\,=\,\frac{L^{4}}{3}z^{-3},
\label{fromztow}
\end{equation}
turns the line element in $(\ref{UGline})$ into the Euclidean AdS metric in Poincar\'e coordinates, namely,
\begin{equation*}
ds^2\,=\,\frac{L^2}{z^2}(dz^2+ \delta_{ij}dx^i dx^j).
\end{equation*}
Since the Riemannian metric coming from the line element in (\ref{UGline}) is unimodular, the coordinates $(w,\vec{x})$ are called unimodular Poincar\'e coordinates. Notice that the boundary of Euclidean AdS is reached when $w\rightarrow\infty$,  for it corresponds to $z=0$.

 The full unimodular metric $\hat{g}_{\mu\nu}$ in (\ref{classaction}) is defined in terms of the background unimodular  metric $\bar{g}_{\mu\nu}$  and the graviton field
 $h_{\mu\nu}$ by the following expressions
\begin{equation}
\hat{g}_{\mu\nu}=\frac{g_{\mu\nu}}{g^{1/4}},\quad\quad g_{\mu\nu}=\bar{g}_{\mu\nu}+\kappa h_{\mu\nu},
\label{hatmetric}
\end{equation}
where $g$ denotes de determinant of $g_{\mu\nu}$.

The symbol $\hat{g}^{(b)}$ in $S_{\text{\tiny{class}}}$ denotes the determinant of the metric induced by $\hat{g}_{\mu\nu}$ on the boundary of Euclidean AdS. $K$ in (\ref{classaction}) is the trace of the extrinsic curvature of the Euclidean AdS boundary for the full unimodular metric $\hat{g}_{\mu\nu}$. Further details can be found in ref. \cite{Anero:2022rqi}

As is well known \cite{Alvarez:2015sba} the theory defined by the action in (\ref{classaction}) is invariant under transverse diffeormophisms and Weyl transformations of the field $g_{\mu\nu}$ in (\ref{hatmetric}).

\newpage
\section{The propagators}
Let us obtain the propagators that we shall need to compute the Witten diagram in Figure 1. The straight lines in that diagram correspond to the boundary-to-bulk propagators of the scalar field with three-momentum $\vec{k}_1$, $\vec{k}_2$, $\vec{k}_3$ and $\vec{k}_4$, respectively. The spiral line denotes the bulk-to-bulk propagator of the graviton.

The boundary-to-bulk propagator, ${\cal G}^{(b)}(w,\vec{x}-\vec{y})$, for the scalar field of our model --see (\ref{classaction})-- is the solution to the homogeneous Laplace  equation on the Euclidean AdS background which is regular when $w\rightarrow 0$ (i.e., $z\rightarrow \infty$):
\begin{equation}
\begin{array}{l}
{\bar{\Box}{\cal G}^{(b)}(w,\vec{x}-\vec{y})=0,}\\[4pt]
{{\cal G}^{(b)}(w,\vec{x}-\vec{y})=\id3k\,{\cal G}^{(b)}(w,\vec{k})\, e^{i\vec{k}\cdot (\vec{x}-\vec{y})},}\\[4pt]
{{\cal G}^{(b)}(w,\vec{k})= G^{(b)}(z=(\frac{L^4}{3 w})^{1/3},\vec{k}),\quad  G^{(b)}(z,\vec{k})=\sqrt{\cfrac{2}{\pi}}\,(k z)^{3/2} K_{3/2}[k z]
}
\label{btBGreen}
\end{array}
\end{equation}
In the previous equation $k=|\vec{k}|$ and $K_{\nu}[u]$ denotes the modified Bessel functions of second kind. Note that $z$ and $w$ are related by (\ref{fromztow}).

\begin{figure}
	\centering
	\begin{tikzpicture}[scale=2.5]
	\draw[]circle(1) (0,0);
	\draw[] (-0.7,0.7)--(-0.3,0);
	\draw[] (-0.7,-0.7)--(-0.3,0);
	\draw[] (0.7,0.7)--(0.3,0);
	\draw[] (0.7,-0.7)--(0.3,0);
	\draw[->]  (-0.5,-0.5)--(-0.3,-0.12);
	\draw[->]  (-0.5,0.5)--(-0.3,0.12);
	\draw[->]  (0.5,0.5)--(0.3,0.12);
	\draw[->]  (0.5,-0.5)--(0.3,-0.12);
	\draw[->]  (-0.2,-0.1)--(0.2,-0.1);
	\draw[spring] (-0.3,0) -- (0.3,0);
	\draw[thick]  (-0.2,-0.1)circle(0pt) node[anchor=north west ] {\tiny{$\vec{k}_1+\vec{k}_2$}};
	\draw[thick]  (-0.42,-0.5)circle(0pt) node[anchor=south west ] {\tiny{$\vec{k}_2$}};
	\draw[thick]  (-0.42,0.5)circle(0pt) node[anchor=north west ] {\tiny{$\vec{k}_1$}};
	\draw[thick]  (0.42,0.5)circle(0pt) node[anchor=north east ] {\tiny{$\vec{k}_3$}};
	\draw[thick]  (0.42,-0.5)circle(0pt) node[anchor=south east ] {\tiny{$\vec{k}_4$}};
	\end{tikzpicture}
	
	\caption{The $s$-channel graviton exchange Witten diagram}\label{fig:WD}
\end{figure}
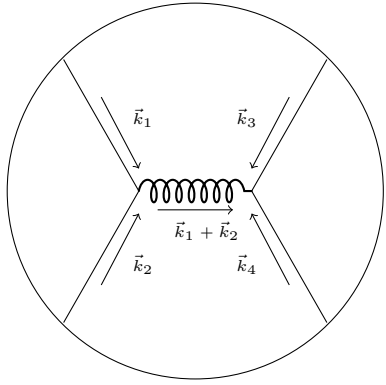

We shall work out next the bulk-to-bulk propagator for the graviton field. We shall carry out this computation in the  axial gauge:
\begin{equation}
h_{0\mu}(w,\vec{x})=0.
\label{axialgauge}
\end{equation}

Let us denote by ${\cal G}_{i_1 j_1,i_2 j_2}(w_1,w_2;\vec{x}_1-\vec{x}_2)$ denote the bulk-to-bulk propagator in the axial gauge. This propagator is by definition the Green's function of the equation of motion of $h_{ij}(w,\vec{x})$ derived from action in (\ref{classaction}) for the axial gauge in (\ref{axialgauge}). The equation of motion in question  reads:
\begin{equation}
 \Box h_{ij}-\frac{3}{8}\bar{g}_{ij}\Box\bar{h}-\bar{\nabla}_i\bar{\nabla}_{\mu} h^{\mu}_j-\bar{\nabla}_j\bar{\nabla}_{\mu} h^{\mu}_i+\frac{1}{2}\bar{g}_{ij}\bar{\nabla}_\mu\bar{\nabla}_\nu h^{\mu \nu}+\frac{1}{2}\bar{\nabla}_i\bar{\nabla}_j \bar{h}+\frac{2}{L^2}h_{ij}-\bar{g}_{ij}\frac{1}{2L^2}\bar{h}=0,
 \label{eomijcov}
 \end{equation}
where $\bar{\nabla}_\mu$ is the covariant derivative for the unimodular background metric $\bar{g}_{\mu\nu}$ in (\ref{BackUGmet}) and $\bar{h}=\bar{g}^{\mu\nu}h_{\mu\nu}$.

By expressing the equation in (\ref{eomijcov}) in terms of the partial derivatives $\partial_w$ and $\partial_{i}$, $i=1,2,3$ one gets
\begin{equation*}
\begin{array}{l}
{\frac{1}{24 L^2 w^{2/3}}\Big((4\sqrt[3]{3}L^{8/3}\partial_i\partial_j +3\delta_{ij}(-\sqrt[3]{3}L^{8/3}\partial^2+6w^{2/3}))h(w,\vec{x})
-8(-\sqrt[3]{3}L^{8/3}\partial^2+6 w^{2/3})h_{ij}(w,\vec{x})}\\[4pt]
{\phantom{\frac{1}{24 L^2 w^{2/3}}\Big(}
+4\sqrt[3]{3}L^{8/3}\delta_{ij}\partial_m\partial_n h_{mn}(w,\vec{x})-8\sqrt[3]{3}L^{8/3}(\partial_i\partial_m h_{mj}(w,\vec{x})+\partial_j\partial_m h_{mi}(w,\vec{x}))}\\[4pt]
{\phantom{\frac{1}{24 L^2 w}\Big(}
-27 w^{5/3}\delta_{ij}(2\partial_w h(w,\vec{x})+3w\partial^2_w h(w,\vec{x}))+72w^{5/3}(2\partial_w h_{ij}(w,\vec{x})+3w\partial^2_w h_{ij}(w,\vec{x}))\Big)=0.
}
\end{array}
\end{equation*}
Bear in mind that in the previous equation $\partial^2\!=\!\delta_{ij}\partial_i\partial_j$ and $h\!=\!\delta_{ij}h_{ij}$.
Then, ${\cal G}_{i_1 j_1,i_2 j_2}(w_1,w_2;\vec{x}_1-\vec{x}_2)$ must satisfy
\begin{equation}
\begin{array}{l}
{\frac{1}{24 L^2 w_1^{2/3}}\Big((4\sqrt[3]{3}L^{8/3}\partial_{i_1}\partial_{j_1} +3\delta_{i_1 j_1}(-\sqrt[3]{3}L^{8/3}\partial^2+6w_1^{2/3})){\cal G}_{mm,i_2 j_2}(w_1,w_2;\vec{x}_1-\vec{x}_2)-}\\[8pt]
{
8(-\sqrt[3]{3}L^{8/3}\partial^2+6 w_1^{2/3}){\cal G}_{i_1 j_1,i_2 j_2}(w_1,w_2;\vec{x}_1-\vec{x}_2)
+4\sqrt[3]{3}L^{8/3}\delta_{i_1j_1}\partial_m\partial_n {\cal G}_{mn,i_2 j_2}(w_1,w_2;\vec{x}_1-\vec{x}_2)-}\\[8pt]
{
8\sqrt[3]{3}L^{8/3}(\partial_{i_1}\partial_m {\cal G}_{m j_1,i_2 j_2}(w_1,w_2;\vec{x}_1-\vec{x}_2)
+\partial_{j_1}\partial_m{\cal G}_{i_1 m,i_2 j_2}(w_1,w_2;\vec{x}_1-\vec{x}_2) )-
}\\[4pt]
{
27 w_1^{5/3}\delta_{i_1 j_1}(2\partial_{w_1}{\cal G}_{mm,i_2 j_2}(w_1,w_2;\vec{x}_1-\vec{x}_2)+3w_1\partial^2_{w_1}{\cal G}_{mm,i_2 j_2}(w_1,w_2;\vec{x}_1-\vec{x}_2))+}\\[4pt]
{
72w_1^{5/3}(2\partial_{w_1}{\cal G}_{i_1 j_1,i_2 j_2}(w_1,w_2;\vec{x}_1-\vec{x}_2)+3w_1\partial^2_{w_1}{\cal G}_{i_1 j_1,i_2 j_2}(w_1,w_2;\vec{x}_1-\vec{x}_2))\Big)=}\\[4pt]
{
\quad\quad\quad\quad\quad\frac{1}{2}(\bar{g}_{i_1 i_2}\bar{g}_{j_1 j_2}+\bar{g}_{i_1 j_2}\bar{g}_{j_1 i_2})\,\delta(w_1-w_2)\,\delta(\vec{x}_1-\vec{x}_2),
}
\label{BtBequationw}
\end{array}
\end{equation}
where $\partial_{i_1}$, $\partial_{j_1}$, $\it{etc}$ denote partial derivatives with regard to $\vec{x}_1$. Recall that $\bar{g}_{\mu\nu}$ has got unit determinant so that no $1/\sqrt{\bar{g}}$ factor occurs on the right hand side of the previous equation. Repeated indices denotes contraction with regard to $\delta_{ij}$.

To solve the equation in (\ref{BtBequationw}) by using the Hankel transform method, one first changes variables from $w$ to $z$ by using (\ref{fromztow}). Upon the change just mentioned, the equation in (\ref{BtBequationw}) becomes
\begin{equation}
\begin{array}{l}
{\frac{1}{8 L^2}\Big(3[-z_1^2\partial^2_{z_1}-2 z_1\partial_{z_1}+(2-z_1^2 \partial^2)]\delta_{i_1 j_1} G^{(UG)}_{m m,i_2 j_2}(z_1,z_2;\vec{x}_1-\vec{x}_2)
+}\\[8pt]
{\phantom{\frac{1}{16 L^2}\Big(}
8[z_1^2\partial^2_{z_1}+2 z_1\partial_{z_1}-(2-z_1^2 \partial^2)] G^{(UG)}_{i_1 j_1,i_2 j_2}(z_1,z_2;\vec{x}_1-\vec{x}_2)+}\\[8pt]
{\phantom{\frac{1}{16 L^2}\Big(}
4 z_1^2\partial_{i_1}\partial_{j_1}G^{(UG)}_{m m,i_2 j_2}(z_1,z_2;\vec{x}_1-\vec{x}_2)+4z_1^2\delta_{i_1 j_1}\partial_{m}\partial_{n} G^{(UG)}_{m n,i_2 j_2}(z_1,z_2;\vec{x}_1-\vec{x_2})-
}\\[8pt]
{\phantom{\frac{1}{16 L^2}\Big(}
8 z_1^2[\partial_{i_1}\partial_{m}G^{(UG)}_{m j_1,i_2 j_2}(z_1,z_2;\vec{x}_1-\vec{x}_2)+\partial_{j_1}\partial_{m}G^{(UG)}_{i_1 m,i_2 j_2}(z_1,z_2;\vec{x}_1-\vec{x}_2)]\Big)=
}\\[8pt]
{\phantom{\frac{1}{16 L^2}\Big(6[-z_1\partial^2_{z_1}}\frac{1}{2}(\delta_{i_1 i_2}\delta_{j_1 j_2}+\delta_{i_1 j_2}\delta_{j_1 i_2})\,\delta(z_1-z_2)\,\delta(\vec{x}_1-\vec{x}_2),
}
\label{BtBeqz}
\end{array}
\end{equation}
where, again, $\partial_{i_1}$, $\partial_{j_1}$, $\it{etc}$ denote partial derivatives with regard to $\vec{x}_1$, repeated indices stands for contraction with regard to $\delta_{ij}$ and
\begin{equation}
G^{(UG)}_{i_1 j_1,i_2 j_2}(z_1,z_2;\vec{x}_1-\vec{x}_2)={\cal G}_{i_1 j_1,i_2 j_2}(w_1=\frac{L^4}{3 z_1^3},w_2=\frac{L^4}{3 z_2^3};\vec{x}_1-\vec{x}_2).
\label{defGUGz1z2}
\end{equation}
Of course, ${\cal G}_{i_1 j_1,i_2 j_2}(w_1,w_2;\vec{x}_1-\vec{x}_2)$ satisfies the equation in (\ref{BtBequationw}).

Now, let us introduce ${\tilde G}^{(UG)}_{i_1 j_1,i_2 j_2}(\omega;\vec{x}_1-\vec{x}_2)$ by means of the Hankel transform:
\begin{equation}
\begin{array}{l}
{
G^{(UG)}_{i_1 j_1,i_2 j_2}(z_1,z_2;\vec{x}_1-\vec{x}_2)=\id3k\,G^{(UG)}_{i_1 j_1,i_2 j_2}(z_1,z_2;\vec{k}) \, e^{i\vec{k}\cdot (\vec{x}-\vec{y})},}\\[8pt]
{
G^{(UG)}_{i_1 j_1,i_2 j_2}(z_1,z_2;\vec{k})=(z_1 z_ 2)^{-1/2}\int_{0}^{\infty} d\omega\,\omega\, J_{3/2}[\omega z_1]\,{\tilde G}^{(UG)}_{i_1 j_1,i_2 j_2}(\omega;\vec{k})\,J_{3/2}[\omega z_2],
}
\label{FBtrans}
\end{array}
\end{equation}
$J_{3/2}[u]$ denotes a Bessel function of first kind.

By substituting the definitions in (\ref{FBtrans}) in the equation in (\ref{BtBeqz}), one ends up with an algebraic equation to be solved by
 ${\tilde G}^{(UG)}_{i_1 j_1,i_2 j_2}(\omega;\vec{k})$. The solution to the algebraic equation in question reads
\begin{equation}
\begin{array}{l}
{{\tilde G}^{(UG)}_{i_1 j_1,i_2 j_2}(\omega;\vec{k})=}\\[8pt]
{
G_{1}^{(UG)}(\vec{k}^2)^2(\delta_{i_1 i_2}\delta_{j_1 j_2}+\delta_{i_1 j_2}\delta_{j_1 i_2})+
G_{2}^{(UG)}(\vec{k}^2)^2\delta_{i_1 j_1}\delta_{i_2 j_2}+
G_{3}^{(UG)}\vec{k}^2(\delta_{i_1 j_1} k_{i_2} k_{j_2}+\delta_{i_2 j_2} k_{i_1} k_{j_1})+}\\[8pt]
{G_{4}^{(UG)}\vec{k}^2(\delta_{i_1 i_2} k_{j_1} k_{j_2}+\delta_{j_1 j_2} k_{i_1} k_{i_2}+
\delta_{i_1 j_2} k_{j_1} k_{i_2}+\delta_{j_1 i_2} k_{i_1} k_{j_2})+
G_{5}^{(UG)} k_{i_1} k_{i_2} k_{i_3} k_{i_4},
}
\label{BtBprop}
\end{array}
\end{equation}
where
\begin{equation}
\begin{array}{l}
{
G_1^{(UG)}=-\cfrac{L^2}{2}\cfrac{1}{(\vec{k}^2)^2(\vec{k}^2+\omega^2)},\quad G_2^{(UG)}=-\cfrac{L^2}{2}\cfrac{1}{(\vec{k}^2)^2(\vec{k}^2+\omega^2)}
\left[\cfrac{2((\vec{k}^2)^2-3\omega^4)}{(\vec{k}^2-\omega^2)^2}\right],
}\\[8pt]
{G_3^{(UG)}=\cfrac{L^2}{2}\cfrac{1}{(\vec{k}^2)^2(\vec{k}^2+\omega^2)}
\left[\cfrac{4 \vec{k}^2\omega^2}{(\vec{k}^2-\omega^2)^2}\right],\quad G_4^{(UG)}=-\cfrac{L^2}{2}\cfrac{1}{(\vec{k}^2)^2(\vec{k}^2+\omega^2)}
\left[\cfrac{\vec{k}^2}{\omega^2}\right],}\\[8pt]
{G_5^{(UG)}=\cfrac{L^2}{2}\cfrac{1}{(\vec{k}^2)^2(\vec{k}^2+\omega^2)}
\left[\cfrac{4(\vec{k}^2)^3}{\omega^2(\vec{k}^2-\omega^2)^2}\right].
}
\label{coefBtBprop}
\end{array}
\end{equation}

Later, well shall also need the bulk-to-bulk propagator  in the axial gauge,$G^{(GR)}_{i_1 j_1,i_2 j_2}(z_1,z_2;\vec{k})$, for General Relativity. The value of this propagator can be found in ref. \cite{Raju:2011mp} and, with the sign convention of ref. \cite{Armstrong:2023phb}, it reads
\begin{equation}
\begin{array}{l}
{
G^{(GR)}_{i_1 j_1,i_2 j_2}(z_1,z_2;\vec{k})=(z_1 z_ 2)^{-1/2}\int_{0}^{\infty} d\omega\,\omega\, J_{3/2}[\omega z_1]\,{\tilde G}^{(GR)}_{i_1 j_1,i_2 j_2}(\omega,\vec{k})\,J_{3/2}[\omega z_2],
}\\[8pt]
{
{\tilde G}^{(GR)}_{i_1 j_1,i_2 j_2}(\omega,\vec{k})= G_1^{(GR)}(T_{i_1 i_2} T_{j_1 j_2}+T_{i_1 j_2} T_{j_1 i_2}-T_{i_1 j_1} T_{i_2 j_2})+
}\\[8pt]
{
\phantom{G^{(GR)}_{i_1 j_1,i_2 j_2}(\omega;\vec{k})=}
G_2^{(GR)}(T_{i_1 i_2} L_{j_1 j_2}+L_{i_1 i_2} T_{j_1 j_2}+T_{i_1 j_2} L_{j_1 i_2}+L_{i_1 j_2} T_{j_1 i_2}  -T_{i_1 j_1} L_{i_2 j_2}-L_{i_1 j_1} T_{i_2 j_2}+
}\\[2pt]
{\phantom{G^{(GR)}_{i_1 j_1,i_2 j_2}(\omega;\vec{k})=G_2^{(GR)}(}
L_{i_1 i_2} L_{j_1 j_2}  +L_{i_1 j_2} L_{j_1 i_2}-L_{i_1 j_1} L_{i_2 j_2})+}\\[8pt]
{
\phantom{G^{(GR)}_{i_1 j_1,i_2 j_2}(\omega;\vec{k})=}
 G_3^{(GR)}(L_{i_1 i_2} L_{j_1 j_2}+L_{i_1 j_2} L_{j_1 i_2}-L_{i_1 j_1} L_{i_2 j_2}),
}
\label{BtBpropGR}
\end{array}
\end{equation}
where
\begin{equation}
\begin{array}{l}
{
T_{ij}=\vec{k}^2 \delta_{ij}-k_i k_j,\quad L_{ij}=k_i k_j},
\\[8pt]
{
G_1^{(GR)}=-\cfrac{L^2}{2}\cfrac{1}{(\vec{k}^2)^2(\vec{k}^2+\omega^2)},
}\\[8pt]
{G_2^{(GR)}=-\cfrac{L^2}{2}\cfrac{1}{(\vec{k}^2)^2(\vec{k}^2+\omega^2)}
\left[\cfrac{\vec{k}^2+\omega^2}{\omega^2}\right],\quad
G_3^{(GR)}=-\cfrac{L^2}{2}\cfrac{1}{(\vec{k}^2)^2(\vec{k}^2+\omega^2)}
\left[\cfrac{\vec{k}^2(\vec{k}^2+\omega^2)}{\omega^4}\right].
}
\label{coefBtBpropGR}
\end{array}
\end{equation}

 Let us now introduce $G^{(re)}_{i_1 j_1,i_2 j_2}(z_1,z_2;\vec{k})$:
\begin{equation}
G^{(UG)}_{i_1 j_1,i_2 j_2}(z_1,z_2;\vec{k})=G^{(GR)}_{i_1 j_1,i_2 j_2}(z_1,z_2;\vec{k})+G^{(re)}_{i_1 j_1,i_2 j_2}(z_1,z_2;\vec{k}).
\label{Gremainderdef}
\end{equation}
Then, by employing (\ref{BtBprop}), (\ref{coefBtBprop}), (\ref{BtBpropGR}) and (\ref{coefBtBpropGR}), one gets
\begin{equation}
\begin{array}{l}
{
G^{(re)}_{i_1 j_1,i_2 j_2}(z_1,z_2;\vec{k})=(z_1 z_ 2)^{-1/2}\int_{0}^{\infty} d\omega\,\omega\, J_{3/2}[\omega z_1]\,{\tilde G}^{(re)}_{i_1 j_1,i_2 j_2}(\omega,\vec{k})\,J_{3/2}[\omega z_2],
}\\[8pt]
{{\tilde G}^{(re)}_{i_1 j_1,i_2 j_2}(\omega;\vec{k})=
G_{2}^{(re)}\delta_{i_1 j_1}\delta_{i_2 j_2}+
G_{3}^{(re)}(\delta_{i_1 j_1} k_{i_2} k_{j_2}+\delta_{i_2 j_2} k_{i_1} k_{j_1})+
G_{5}^{(re)} k_{i_1} k_{i_2} k_{i_3} k_{i_4},
}\\[8pt]
{
G_2^{(re)}=\cfrac{L^2}{2}\cfrac{1}{(\vec{k}^2)^2(\vec{k}^2+\omega^2)}
\left[\cfrac{-3(\vec{k}^2)^4+2(\vec{k}^2)^3 \omega^2+5(\vec{k}^2)^2\omega^4}{(\vec{k}^2-\omega^2)^2}\right],\quad
G_3^{(re)}=\cfrac{L^2}{2}\left[\cfrac{-\vec{k}^2+3\omega^2}{\omega^2(\vec{k}^2-\omega^2)^2}\right],
}\\[8pt]
{
G_5^{(re)}=\cfrac{L^2}{2}\left[\cfrac{\vec{k}^2+\omega^2}{\omega^4(\vec{k}^2-\omega^2)^2}\right].
}
\label{reprop}
\end{array}
\end{equation}

Before we close the current section, a comment regarding the pole structure of ${\tilde G}^{(UG)}_{i_1 j_1,i_2 j_2}(\omega;\vec{k})$ in (\ref{BtBprop}) is in order. First, notice that if we compare ${\tilde G}^{(UG)}_{i_1 j_1,i_2 j_2}(\omega;\vec{k})$ with the corresponding object in General Relativity --in (\ref{BtBpropGR}), we will come to the conclusion that a new type of poles arise:  poles at $\omega^2=\vec{k}^2$. We shall move those poles to the complex plane by introducing a small positive imaginary part:
\begin{equation*}
\cfrac{1}{(\vec{k}^2-\omega^2)^2}\equiv\lim_{\epsilon\rightarrow 0^{+}}\cfrac{1}{(\vec{k}^2-\omega^2-i\epsilon)^2}=\lim_{\epsilon\rightarrow 0^{+}}\cfrac{1}{((|\vec{k}|-i\epsilon)^2-\omega^2)^2}
\end{equation*}
This way of going around the poles is obtained by Wick rotation of $k^0$ of the  corresponding expression for $\text{AdS}_4$. Indeed, it can be seen that, for $\text{AdS}_4$,  we have
\begin{equation}
\begin{array}{l}
{
\cfrac{1}{(-(k^0)^2\!+\!(k^1)^2\!+\!(k^2)^2-\omega^2)^2}\equiv\lim_{\epsilon\rightarrow 0^{+}}\cfrac{1}{(-(k^0)^2\!+\sum_{i=1}^2\!(k^i)^2\!-\omega^2-i\epsilon)^2}=}\\[8pt]
{
\lim_{\epsilon\rightarrow 0^{+}}\cfrac{1}{(-(k^0)^2+(\Omega-i\epsilon)^2)^2},
}
\label{wickrot}
\end{array}
\end{equation}
where we take $\Omega=\sqrt{(k^1)^2+(k^2)^2-\omega^2}\geq 0$, for there is no pole if $(k^1)^2+(k^2)^2-\omega^2\leq 0$. Wick rotation --i.e., $k^0\rightarrow ik^3$-- of (\ref{wickrot}) is allowed, for the poles in $k^0$ occur in the second and fourth quadrant of complex $k^0$-plane. The Wick rotation of the distribution in (\ref{wickrot}) yields
\begin{equation*}
\begin{array}{l}
{\lim_{\epsilon\rightarrow 0^{+}}\cfrac{1}{(-(k^0)^2+(\Omega-i\epsilon)^2)^2}\longrightarrow}\\[8pt]
{\lim_{\epsilon\rightarrow 0^{+}}\cfrac{1}{((k^3)^2+(\Omega-i\epsilon)^2)^2}=
\lim_{\epsilon\rightarrow 0^{+}}\cfrac{1}{((k^3)^2+\Omega^2-i\epsilon)^2}=\lim_{\epsilon\rightarrow 0^{+}}\cfrac{1}{(\vec{k}^2-\omega^2-i\epsilon)^2}.
}
\end{array}
\end{equation*}
The reader should bare in mind that all the limits above are to be understood in the sense of theory of distributions.

\newpage

\section{Graviton exchange in the s-channel}

The lowest order interaction term  between the graviton field $h_{\mu\nu}$ and the scalar field $\phi$ in $S_{class}$ in (\ref{classaction}) reads
\begin{equation}
-\frac{\kappa}{2}\int_{0}^\infty\! dw\,\int\! d^{3}\vec{x}\,(\bar{g}^{\mu\rho}\bar{g}^{\nu\sigma}
-\frac{1}{4}\bar{g}^{\mu\nu}\bar{g}^{\rho\sigma})\,\partial_{\rho}\phi(w,\vec{x})\partial_{\sigma}\phi(w,\vec{x})\, h_{\mu\nu}(w,\vec{x}),
\label{interactiont}
\end{equation}
where $\bar{g}^{\mu\nu}$ is the inverse of $\bar{g}_{\mu\nu}$ in (\ref{BackUGmet}). Of course, the previous equation just tells us that in unimodular gravity the graviton field only couples to the traceless part of the energy-momentum tensor. This is unlike in General Relativity where the graviton field couples to the full energy-momentum tensor yielding the following interaction term
\begin{equation*}
-\frac{\kappa}{2}\int_{0}^\infty\! dw\,\int\! d^{3}\vec{x}\,(\bar{g}^{\mu\rho}\bar{g}^{\nu\sigma}
-\frac{1}{2}\bar{g}^{\mu\nu}\bar{g}^{\rho\sigma})\,\partial_{\rho}\phi(w,\vec{x})\partial_{\sigma}\phi(w,\vec{x})\, h_{\mu\nu}(w,\vec{x}).
\end{equation*}

Let us draw next the reader's attention to the boundary-to-bulk propagator in momentum space, given in (\ref{btBGreen}),  and to the axial-gauge bulk-to-bulk propagator,  which has been introduced in the paragraph right below (\ref{axialgauge}) and also displayed in  (\ref{defGUGz1z2}) and (\ref{FBtrans}). The reader should also familiarize with (\ref{BtBprop}) and (\ref{coefBtBprop}). We are now ready to display the mathematical object corresponding to the momentum-space Witten diagram in Figure 1. The object in question yields the s-channel exchange of a graviton in the axial gauge and reads
\begin{equation}
{\cal W}^{(UG)}_s\!=\!(2\pi)^3\delta\big(\sum_{a=1}^4\vec{k}_a\big)\kappa^2\int_{0}^\infty\!\! dw_1\int_{0}^\infty\!\! dw_2\; T_L^{i_1 j_1}(w_1;k_1,k_2)\,{\cal G}_{i_1 j_1,i_2 j_2}
(w_1,w_ 2;\vec{k})\, T_R^{i_2 j_2}(w_2;k_3,k_4),
\label{Wsw1w2}
\end{equation}
where $\vec{k}=\vec{k}_1+\vec{k}_2=-(\vec{k}_3+\vec{k}_4)$ and
\begin{equation*}
\begin{array}{l}
{
{\cal G}_{i_1 j_1,i_2 j_2}
(w_1,w_ 2;\vec{k})=G^{(UG)}_{i_1 j_1,i_2 j_2}(z_1=\sqrt[3]{L^4/ (3 w_1)},z_ 2=\sqrt[3]{L^4/ (3 w_2)} ;\vec{k}),}\\[8pt]
{
T_L^{i_1 j_1}(w_1;k_1,k_2)=
-\bar{g}^{i_1 m_1}\bar{g}^{j_1 n_1}k_{1 m_1} k_{2 n_1}{\cal G}^{(b)}(w_1,\vec{k}_1){\cal G}^{(b)}(w_1,\vec{k}_2)-}\\[8pt]
{\phantom{T_L^{i_1 j_1}(w_1;k_1}
\frac{1}{4}\bar{g}^{i_1 j_1}
[\bar{g}^{w_1 w_1}\partial_{w_1}{\cal G}^{(b)}(w_1,\vec{k}_1)\partial_{w_1}{\cal G}^{(b)}(w_1,\vec{k}_2)-\bar{g}^{nm}k_{1m} k_{2n}\,{\cal G}^{(b)}(w_1,\vec{k}_1){\cal G}^{(b)}(w_1,\vec{k}_2)],
}\\[8pt]
{
T_R^{i_2 j_2}(w_2;k_3,k_4)=
-\bar{g}^{i_2 m_2}\bar{g}^{j_2 n_2}k_{3 m_2} k_{3 n_2}{\cal G}^{(b)}(w_2,\vec{k}_3){\cal G}^{(b)}(w_2,\vec{k}_4)-}\\[8pt]
{\phantom{T_L^{i_1 j_1}(w_1;k_1}
\frac{1}{4}\bar{g}^{i_2 j_2}
[\bar{g}^{w_2 w_2}\partial_{w_2}{\cal G}^{(b)}(w_2,\vec{k}_3)\partial_{w_2}{\cal G}^{(b)}(w_2,\vec{k}_4)-\bar{g}^{nm}k_{3m} k_{4n}\,{\cal G}^{(b)}(w_2,\vec{k}_3){\cal G}^{(b)}(w_2,\vec{k}_4)].
}
\end{array}
\end{equation*}
Obviously, $T_L^{i_1 j_1}(w_1;k_1,k_2)$ and $T_R^{i_2 j_2}(w_2;k_3,k_4)$ come from the interaction term in (\ref{interactiont}). The background metric $\bar{g}_{\mu\nu}$ is displayed in (\ref{BackUGmet}).

The change of variables
\begin{equation*}
w_1=\frac{L^4}{3 z_1^3},\quad w_2=\frac{L^4}{3 z_2^3},
\end{equation*}
simplifies the expression defining ${\cal W}^{(UG)}_s$ in (\ref{Wsw1w2}). Indeed, we have
\begin{equation*}
{\cal W}^{(UG)}_s\!=\!(2\pi)^3\delta\big(\sum_{a=1}^4\,\vec{k}_a\big)\kappa^2\!\!\int_{0}^\infty\!\!\! dz_1\!\!\int_{0}^\infty\!\!\! dz_2 V_L^{(UG)\,i_1 j_1}(z_1;k_1,\!k_2)G^{(UG)}_{i_1 j_1,i_2 j_2}
(z_1,\!z_ 2;\vec{k}) V_R^{(UG)\,i_2 j_2}(z_2;k_3,\!k_4),
\end{equation*}
where $G^{(UG)}_{i_1 j_1,i_2 j_2}(z_1,z_ 2;\vec{k})$ is given in (\ref{FBtrans}), (\ref{BtBprop}) and (\ref{coefBtBprop}) and
\begin{equation*}
\begin{array}{l}
{
V_L^{(UG)\,i_1 j_1}(z_1;k_1,k_2)=
-\delta^{i_1 m_1}\delta^{j_1 n_1}k_{1 m_1} k_{2 n_1}G^{(b)}(z_1,\vec{k}_1)G^{(b)}(z_1,\vec{k}_2)-}\\[8pt]
{\phantom{V_L^{(UG)\,i_1 j_1}(w_1;k_1,k_2)=}
\frac{1}{4}\delta^{i_1 j_1}
[\partial_{z_1}G^{(b)}(z_1,\vec{k}_1)\partial_{z_1}G^{(b)}(z_1,\vec{k}_2)-\vec{k}_1\cdot\vec{k}_2\,G^{(b)}(z_1,\vec{k}_1)G^{(b)}(z_1,\vec{k}_2)],
}\\[8pt]
{
V_R^{(UG)\,i_2 j_2}(z_2;k_3,k_4)=
-\delta^{i_2 m_2}\delta^{j_2 n_2}k_{3 m_2} k_{3 n_2}G^{(b)}(z_2,\vec{k}_3)G^{(b)}(z_2,\vec{k}_4)-}\\[8pt]
{\phantom{V_L^{(UG)\,i_1 j_1}(z_1;k_1,k_2)=}
\frac{1}{4}\delta^{i_2 j_2}
[\partial_{z_2}G^{(b)}(z_2,\vec{k}_3)\partial_{z_2} G^{(b)}(z_2,\vec{k}_4)-\vec{k}_3\cdot\vec{k}_4\, G^{(b)}(z_2,\vec{k}_3)G^{(b)}(z_2,\vec{k}_4)].
}
\end{array}
\end{equation*}
$G^{(b)}(z_2,\vec{k}_4)$ is defined in (\ref{btBGreen}).

The main purpose of this paper is to compute ${\cal W}^{(UG)}_s$ above and compare it with the corresponding General Relativity quantity $W^{(GR)}_s$. To do so we shall compute first the difference between the ${\cal W}^{(UG)}_s$ and $W^{(GR)}_s$. It is plane that
\begin{equation}
{\cal W}^{(GR)}_s\!\!=\!\!(2\pi)^3\!\delta\big(\sum_{a=1}^4\vec{k}_a\big)\kappa^2\!\!\!\int_{0}^\infty\!\!\! dz_1\!\!\int_{0}^\infty\!\!\! dz_2\; V_L^{(GR)\,i_1 j_1}(z_1;k_1,k_2)\,G^{(GR)}_{i_1 j_1,i_2 j_2}
(z_1,z_ 2;\vec{k})\, V_R^{(GR)\,i_2 j_2}(z_2;k_3,k_4),
\label{WGRsz1z2}
\end{equation}
where $G^{(GR)}_{i_1 j_2,i_2 j_2}(z_1,z_ 2;\vec{k})$ is to be found in (\ref{BtBpropGR}) and
\begin{equation*}
\begin{array}{l}
{
V_L^{(GR)\,i_1 j_1}(z_1;k_1,k_2)=
-\delta^{i_1 m_1}\delta^{j_1 n_1}k_{1 m_1} k_{2 n_1}G^{(b)}(z_1,\vec{k}_1)G^{(b)}(z_1,\vec{k}_2)-}\\[4pt]
{\phantom{V_L^{(GR)\,i_1 j_1}(z_1;k_1,k_2)=}
\frac{1}{2}\delta^{i_1 j_1}
[\partial_{z_1}G^{(b)}(z_1,\vec{k}_1)\partial_{z_1}G^{(b)}(z_1,\vec{k}_2)-\vec{k}_1\cdot\vec{k}_2\,G^{(b)}(z_1,\vec{k}_1)G^{(b)}(z_1,\vec{k}_2)],
}\\[8pt]
{
V_R^{(GR)\,i_2 j_2}(z_2;k_3,k_4)=
-\delta^{i_2 m_2}\delta^{j_2 n_2}k_{3 m_2} k_{3 n_2}G^{(b)}(z_2,\vec{k}_3)G^{(b)}(z_2,\vec{k}_4)-}\\[4pt]
{\phantom{V_L^{(GR)\,i_1 j_1}(z_1;k_1,k_2)=}
\frac{1}{2}\delta^{i_2 j_2}
[\partial_{z_2}G^{(b)}(z_2,\vec{k}_3)\partial_{z_2} G^{(b)}(z_2,\vec{k}_4)-\vec{k}_3\cdot\vec{k}_4\, G^{(b)}(z_2,\vec{k}_3)G^{(b)}(z_2,\vec{k}_4)].
}
\end{array}
\end{equation*}

Using the splitting in (\ref{Gremainderdef}) and the fact that
\begin{equation*}
\begin{array}{l}
{V_L^{(UG)\,i_1 j_1}(z_1;k_1,k_2)=V_L^{(GR)\,i_1 j_1}(z_1;k_1,k_2)+V_L^{(re)\,i_1 j_1}(z_1;k_1,k_2),}\\[2pt]
{V_R^{(UG)\,i_2 j_2}(z_2;k_3,k_4)=V_R^{(GR)\,i_2 j_2}(z_2;k_3,k_4)+V_R^{(re)\,i_2 j_2}(z_2;k_3,k_4),}
\end{array}
\end{equation*}
if
\begin{equation*}
\begin{array}{l}
{
V_L^{(re)\,i_1 j_1}(z_1;k_1,k_2)=
\frac{1}{4}\delta^{i_1 j_1}
[\partial_{z_1}G^{(b)}(z_1,\vec{k}_1)\partial_{z_1}G^{(b)}(z_1,\vec{k}_2)-\vec{k}_1\cdot\vec{k}_2\,G^{(b)}(z_1,\vec{k}_1)G^{(b)}(z_1,\vec{k}_2)],
}\\[4pt]
{
V_R^{(re)\,i_2 j_2}(z_2;k_3,k_4)=
\frac{1}{4}\delta^{i_2 j_2}
[\partial_{z_2}G^{(b)}(z_2,\vec{k}_3)\partial_{z_2} G^{(b)}(z_2,\vec{k}_4)-\vec{k}_3\cdot\vec{k}_4\, G^{(b)}(z_2,\vec{k}_3)G^{(b)}(z_2,\vec{k}_4)],
}
\end{array}
\end{equation*}
one concludes that
\begin{equation}
{\cal W}^{(UG)}_s={\cal W}^{(GR)}_s+{\cal W}^{(re)}_s,\quad\quad {\cal W}^{(re)}_s=(2\pi)^3\delta\big(\sum_{a=1}^4\,\vec{k}_a\big)\,(\text{C}_1+\text{C}_2+\text{C}_3+\text{C}_4),
\label{theWs}
\end{equation}
where
\begin{equation}
\begin{array}{l}
{
\text{C}_1=\kappa^2\int_{0}^\infty\! dz_1\int_{0}^\infty\! dz_2\; V_L^{(GR)\,i_1 j_1}(z_1;k_1,k_2)\,G^{(re)}_{i_1 j_1,i_2 j_2}
(z_1,z_ 2;\vec{k})\, V_R^{(GR)\,i_2 j_2}(z_2;k_3,k_4),
}\\[2pt]
{
\text{C}_2=\kappa^2\int_{0}^\infty\! dz_1\int_{0}^\infty\! dz_2\; V_L^{(GR)\,i_1 j_1}(z_1;k_1,k_2)\,G^{(UG)}_{i_1 j_1,i_2 j_2}
(z_1,z_ 2;\vec{k})\, V_R^{(re)\,i_2 j_2}(z_2;k_3,k_4),
}\\[2pt]
{
\text{C}_3=\kappa^2\int_{0}^\infty\! dz_1\int_{0}^\infty\! dz_2\; V_L^{(re)\,i_1 j_1}(z_1;k_1,k_2)\,G^{(UG)}_{i_1 j_1,i_2 j_2}
(z_1,z_ 2;\vec{k})\, V_R^{(GR)\,i_2 j_2}(z_2;k_3,k_4),
}\\[2pt]
{
\text{C}_4=\kappa^2\int_{0}^\infty\! dz_1\int_{0}^\infty\! dz_2\; V_L^{(re)\,i_1 j_1}(z_1;k_1,k_2)\,G^{(UG)}_{i_1 j_1,i_2 j_2}
(z_1,z_ 2;\vec{k})\, V_R^{(re)\,i_2 j_2}(z_2;k_3,k_4).
}
\label{C1234coeff}
\end{array}
\end{equation}
Let us note that the head-on computation of $\text{C}_1$, $\text{C}_2$, $\text{C}_3$ and $\text{C}_4$ involves the calculation of 32 integrals over the variables $z_1$, $z_2$ and $\omega$ whose integrands contain Bessel functions. The value of these integrals can be found in Appendix A and  they will be denoted by the symbols $\text{I}_{ab}$, where  $a=1,2,3,4$ and $b=1,2,3,4,5,6,7,8$.

To to work out the values of $\text{C}_1$, $\text{C}_2$, $\text{C}_3$ and $\text{C}_4$ in (\ref{C1234coeff}), we have carried out first the vector algebra involving  $\delta_{ij}$ and $k_i$. The contractions that this computation involve have done by using the symbolic manipulation system FORM \cite{Ruijl:2017dtg}. Thus we have obtained the following results:
\begin{equation}
\begin{array}{l}
{
\text{C}_1=\frac{L^2}{16}\,\big\{36 \,\text{I}_{16} + 6 \,\text{I}_{36} \,k_1^2 - 6 \,\text{I}_{37} \,k_1^4 + 6 \,\text{I}_{36} \,k_2^2 +
   12 \,\text{I}_{37} \,k_1^2 \,k_2^2 - 6 \,\text{I}_{37} \,k_2^4 +
    \,\text{I}_{46} \,k_1^2 \,k_3^2 - \,\text{I}_{47} \,k_1^4 \,k_3^2 +}\\[2pt]
 {\phantom{\text{C}}
   \,\text{I}_{46} \,k_2^2 \,k_3^2 + 2 \,\text{I}_{47} \,k_1^2 \,k_2^2 \,k_3^2 - \,\text{I}_{47} \,k_2^4 \,k_3^2 - 6 \,\text{I}_{27} \,k_3^4 -
    \,\text{I}_{47} \,k_1^2 \,k_3^4 + \,\text{I}_{48} \,k_1^4 \,k_3^4 - \,\text{I}_{47} \,k_2^2 \,k_3^4 -}\\[2pt]
{\phantom{\text{C}}
     2 \,\text{I}_{48} \,k_1^2 \,k_2^2 \,k_3^4 +
   \,\text{I}_{48} \,k_2^4 \,k_3^4 + \,\text{I}_{46} \,k_1^2 \,k_4^2 -
   \,\text{I}_{47} \,k_1^4 \,k_4^2 + \,\text{I}_{46} \,k_2^2 \,k_4^2 + 2 \,\text{I}_{47} \,k_1^2 \,k_2^2 \,k_4^2 -
   \,\text{I}_{47} \,k_2^4 \,k_4^2 + }\\[2pt]
   {\phantom{\text{C}}
   12 \,\text{I}_{27} \,k_3^2 \,k_4^2 + 2 \,\text{I}_{47} \,k_1^2 \,k_3^2 \,k_4^2 -
   2 \,\text{I}_{48} \,k_1^4 \,k_3^2 \,k_4^2 + 2 \,\text{I}_{47} \,k_2^2 \,k_3^2 \,k_4^2 +
   4 \,\text{I}_{48} \,k_1^2 \,k_2^2 \,k_3^2 \,k_4^2 -
    2 \,\text{I}_{48} \,k_2^4 \,k_3^2 \,k_4^2 -}\\[2pt]
{\phantom{\text{C}}
     6 \,\text{I}_{27} \,k_4^4 -
   \,\text{I}_{47} \,k_1^2 \,k_4^4 + \,\text{I}_{48} \,k_1^4 \,k_4^4 - \,\text{I}_{47} \,k_2^2 \,k_4^4 - 2 \,\text{I}_{48} \,k_1^2 \,k_2^2 \,k_4^4 +
\,\text{I}_{48} \,k_2^4 \,k_4^4 + 6 \,\text{I}_{26} (\,k_3^2 + \,k_4^2 - \,s) +}\\[2pt]
 {\phantom{\text{C}}
   24 \,\text{I}_{17} \,s - (6 \,\text{I}_{36} + \,\text{I}_{46} \,k_1^2 + 2 \,\text{I}_{38} \,k_1^4 - \,\text{I}_{47} \,k_1^4 + \,\text{I}_{46} \,k_2^2 -
      4 \,\text{I}_{38} \,k_1^2 \,k_2^2 + 2 \,\text{I}_{47} \,k_1^2 \,k_2^2 + 2 \,\text{I}_{38} \,k_2^4 -}\\[2pt]
{\phantom{\text{C}}
       \,\text{I}_{47} \,k_2^4 -
      8 \,\text{I}_{37} (\,k_1^2 + \,k_2^2) - 8 \,\text{I}_{27} \,k_3^2 + \,\text{I}_{46} \,k_3^2 -
       2 \,\text{I}_{47} \,k_1^2 \,k_3^2 +
      \,\text{I}_{48} \,k_1^4 \,k_3^2 - 2 \,\text{I}_{47} \,k_2^2 \,k_3^2 -}\\[2pt]
{\phantom{\text{C}}
      2 \,\text{I}_{48} \,k_1^2 \,k_2^2 \,k_3^2 +
      \,\text{I}_{48} \,k_2^4 \,k_3^2 +
       2 \,\text{I}_{28} \,k_3^4 - \,\text{I}_{47} \,k_3^4 + \,\text{I}_{48} \,k_1^2 \,k_3^4 +
      \,\text{I}_{48} \,k_2^2 \,k_3^4 + (-8 \,\text{I}_{27} + \,\text{I}_{46} +}\\[2pt]
{\phantom{\text{C}}
       \,\text{I}_{48} (\,k_1^2 - \,k_2^2)^2 -
         2 (2 \,\text{I}_{28} + \,\text{I}_{48} (\,k_1^2 + \,k_2^2)) \,k_3^2 -
         2 \,\text{I}_{47} (\,k_1^2 + \,k_2^2 - \,k_3^2)) \,k_4^2 + (2 \,\text{I}_{28} - \,\text{I}_{47} +}\\[2pt]
{\phantom{\text{C}}
         \,\text{I}_{48} (\,k_1^2 + \,k_2^2)) \,k_4^4) \,s +
  (4 \,\text{I}_{18} - 2 \,\text{I}_{27} - 2 \,\text{I}_{37} + \,\text{I}_{46} +
      2 \,\text{I}_{38} (\,k_1^2 + \,k_2^2) + (2 \,\text{I}_{28} + \,\text{I}_{48} (\,k_1^2 + \,k_2^2))}\\[2pt]
 {\phantom{\text{C}}
       (\,k_3^2 +\,k_4^2) - \,\text{I}_{47} (\,k_1^2 + \,k_2^2 + \,k_3^2 + \,k_4^2)) \,s^2)\big\}
      }
\label{C1coeff}
\end{array}
\end{equation}
and
\begin{equation}
\begin{array}{l}
{C_2+C_3+C_4=\frac{L^2}{64}\,s\,\big\{}\\[4pt]
{\phantom{C}
4\, \text{I}_{35} \,k_1^4 - 8 \,\text{I}_{35} \,k_1^2 \,k_2^2 + 4 \,\text{I}_{35} \,k_2^4 + 6 \,\text{I}_{43} \,k_1^4 \,k_3^2 +
   8 \,\text{I}_{44} \,k_1^4 \,k_3^2 + 2 \,\text{I}_{45} \,k_1^4 \,k_3^2 -
    12 \,\text{I}_{43} \,k_1^2 \,k_2^2 \,k_3^2 -}\\[2pt]
{\phantom{C}
   16 \,\text{I}_{44} \,k_1^2 \,k_2^2 \,k_3^2 - 4 \,\text{I}_{45} \,k_1^2 \,k_2^2 \,k_3^2 + 6 \,\text{I}_{43} \,k_2^4 \,k_3^2 +
   8 \,\text{I}_{44} \,k_2^4 \,k_3^2 + 2 \,\text{I}_{45} \,k_2^4 \,k_3^2 + 12 \,\text{I}_{23} \,k_3^4 + 16 \,\text{I}_{24} \,k_3^4 +}\\[2pt]
{\phantom{C}
   4 \,\text{I}_{25} \,k_3^4 + 6 \,\text{I}_{43} \,k_1^2 \,k_3^4 + 8 \,\text{I}_{44} \,k_1^2 \,k_3^4 + 2 \,\text{I}_{45} \,k_1^2 \,k_3^4 +
   6 \,\text{I}_{43} \,k_2^2 \,k_3^4 + 8 \,\text{I}_{44} \,k_2^2 \,k_3^4 + 2 \,\text{I}_{45} \,k_2^2 \,k_3^4 +}\\[2pt]
{\phantom{C}
   6 \,\text{I}_{43} \,k_1^4 \,k_4^2 + 8 \,\text{I}_{44} \,k_1^4 \,k_4^2 + 2 \,\text{I}_{45} \,k_1^4 \,k_4^2 -
   12 \,\text{I}_{43} \,k_1^2 \,k_2^2 \,k_4^2 - 16 \,\text{I}_{44} \,k_1^2 \,k_2^2 \,k_4^2 -
   4 \,\text{I}_{45} \,k_1^2 \,k_2^2 \,k_4^2 +}\\[2pt]
{\phantom{C}
   6 \,\text{I}_{43} \,k_2^4 \,k_4^2 + 8 \,\text{I}_{44} \,k_2^4 \,k_4^2 +
   2 \,\text{I}_{45} \,k_2^4 \,k_4^2\! -\! 24 \,\text{I}_{23} \,k_3^2 \,k_4^2\! -\! 32 \,\text{I}_{24} \,k_3^2 \,k_4^2\! -\!
   8 \,\text{I}_{25} \,k_3^2 \,k_4^2\! -\! 12 \,\text{I}_{43} \,k_1^2 \,k_3^2 \,k_4^2\! -}\\[2pt]
{\phantom{C}
    16 \,\text{I}_{44} \,k_1^2 \,k_3^2 \,k_4^2 -
   4 \,\text{I}_{45} \,k_1^2 \,k_3^2 \,k_4^2 - 12 \,\text{I}_{43} \,k_2^2 \,k_3^2 \,k_4^2 -
   16 \,\text{I}_{44} \,k_2^2 \,k_3^2 \,k_4^2 - 4 \,\text{I}_{45} \,k_2^2 \,k_3^2 \,k_4^2 + 12\,\text{I}_{23} \,k_4^4 +}\\[2pt]
{\phantom{C}
   16 \,\text{I}_{24} \,k_4^4 + 4 \,\text{I}_{25} \,k_4^4 + 6 \,\text{I}_{43} \,k_1^2 \,k_4^4 + 8 \,\text{I}_{44} \,k_1^2 \,k_4^4 +
   2 \,\text{I}_{45} \,k_1^2 \,k_4^4 + 6 \,\text{I}_{43} \,k_2^2 \,k_4^4 + 8 \,\text{I}_{44} \,k_2^2 \,k_4^4 +}\\[2pt]
{\phantom{C}
   2 \,\text{I}_{45} \,k_2^2 \,k_4^4 -
   12 (6 \,\text{I}_{11} + 9 \,\text{I}_{12} + 6 \,\text{I}_{13} + 4 \,\text{I}_{14} + \,\text{I}_{15}) \,s - (6 \,\text{I}_{35} \,k_1^2 +
      6 \,\text{I}_{43} \,k_1^4 + 8 \,\text{I}_{44} \,k_1^4 + 2 \,\text{I}_{45} \,k_1^4 +}\\[2pt]
{\phantom{C}
       6 \,\text{I}_{35} \,k_2^2 -
      12 \,\text{I}_{43} \,k_1^2 \,k_2^2 - 16 \,\text{I}_{44} \,k_1^2 \,k_2^2 - 4 \,\text{I}_{45} \,k_1^2 \,k_2^2 +
      6 \,\text{I}_{43} \,k_2^4 + 8 \,\text{I}_{44} \,k_2^4 + 2 \,\text{I}_{45} \,k_2^4 +}\\[2pt]
{\phantom{C}
       20 \,\text{I}_{31} (\,k_1^2 + \,k_2^2) +
      30 \,\text{I}_{32} (\,k_1^2 + \,k_2^2) + 2 \,\text{I}_{41} \,k_1^2 \,k_3^2 + 3 \,\text{I}_{42} \,k_1^2 \,k_3^2 +
      10 \,\text{I}_{43} \,k_1^2 \,k_3^2 + 12 \,\text{I}_{44} \,k_1^2 \,k_3^2 +}\\[2pt]
{\phantom{C}
       3 \,\text{I}_{45} \,k_1^2 \,k_3^2 +
      2 \,\text{I}_{41} \,k_2^2 \,k_3^2 + 3 \,\text{I}_{42} \,k_2^2 \,k_3^2 + 10 \,\text{I}_{43} \,k_2^2 \,k_3^2 +
      12 \,\text{I}_{44} \,k_2^2 \,k_3^2 + 3 \,\text{I}_{45} \,k_2^2 \,k_3^2 + 6 \,\text{I}_{43} \,k_3^4 +}\\[2pt]
 {\phantom{C}
      8 \,\text{I}_{44} \,k_3^4\! +\!
      2 \,\text{I}_{45} \,k_3^4\! +\! ((2 \,\text{I}_{41}\! +\! 3 \,\text{I}_{42}\! +\! 10 \,\text{I}_{43}\! +\! 12 \,\text{I}_{44}\! +\! 3 \,\text{I}_{45}) (\,k_1^2\! +\!
            \,k_2^2) - 4 (3 \,\text{I}_{43}\! +\! 4 \,\text{I}_{44}\! +\! \,\text{I}_{45}) \,k_3^2) \,k_4^2\! +}\\[2pt]
 {\phantom{C}
      2 (3 \,\text{I}_{43} + 4 \,\text{I}_{44} + \,\text{I}_{45}) \,k_4^4 +
      2 (10 \,\text{I}_{21} + 15 \,\text{I}_{22} + 14 \,\text{I}_{23} + 12 \,\text{I}_{24} + 3 \,\text{I}_{25}) (\,k_3^2 +
         \,k_4^2)) \,s + (20 \,\text{I}_{21} +}\\[2pt]
{\phantom{C}
          30 \,\text{I}_{22} + 16 \,\text{I}_{23} + 8 \,\text{I}_{24} + 2 \,\text{I}_{25} +
      20 \,\text{I}_{31} + 30 \,\text{I}_{32} +
      2 \,\text{I}_{35} + (2 \,\text{I}_{41} + 3 \,\text{I}_{42} + 4 (\,\text{I}_{43} + \,\text{I}_{44}) + \,\text{I}_{45})}\\[2pt]
{\phantom{C}
       (\,k_1^2 + \,k_2^2 +
         \,k_3^2 + \,k_4^2)) \,s^2 + (-2 \,\text{I}_{41} - 3 \,\text{I}_{42} + 2 \,\text{I}_{43} + 4 \,\text{I}_{44} +
      \,\text{I}_{45}) \,s^3 +
   8 \,\text{I}_{34} (2 (\,k_1^2 - \,k_2^2)^2 -}\\[2pt]
{\phantom{C}
    3 (\,k_1^2 + \,k_2^2) \,s + \,s^2) +
   4 \,\text{I}_{33} (3 (\,k_1^2 - \,k_2^2)^2 - 7 (\,k_1^2 + \,k_2^2) \,s + 4 \,s^2)\big\},
}
\label{C234coeff}
\end{array}
\end{equation}
where $k_a$, $a=1,2,3$, is the modulus of the vector $\vec{k}_a$ and $s$ is Mandelstam variable $s=(\vec{k_1}+\vec{k_2})^2$. Recall that $\vec{k}_1+\vec{k}_2+\vec{k}_3+\vec{k}_4=0$.

To obtain the value of ${\cal W}^{(re)}_s$ in (\ref{theWs}), the integrals in Appendix A are to be substituted in the expressions for $\text{C}_1$, in (\ref{C1coeff}), and $\text{C}_{2}+\text{C}_3+\text{C}_{4}$, in (\ref{C234coeff}), and then add the results. This we have done by using Mathematica \cite{mathematica} and obtained a vanishing  result:
\begin{equation*}
{\cal W}^{(re)}_s=(2\pi)^3\delta\big(\sum_{a=1}^4\,\vec{k}_a\big)\,(\text{C}_1+\text{C}_2+\text{C}_3+\text{C}_4)=0,\quad \forall\, \vec{k}_a,\;a=1,2,3,4.
\end{equation*}
Hence, and according to (\ref{theWs}), one concludes that the value of the $s$-channel Witten diagram in Figure 1 is the same in unimodular gravity as in General Relativity:
\begin{equation*}
{\cal W}^{(UG)}_s={\cal W}^{(GR)}_s.
\end{equation*}
Of course, the same result holds for the Witten diagrams corresponding to the $t=(\vec{k_1}+\vec{k_3})^2$ and $u=(\vec{k_1}+\vec{k_4})^2$ channels.

We have also computed anew  the value of the $s$-channel Witten diagram in Figure 1 --${\cal W}^{(GR)}_s$-- for General Relativity. But before giving the result of our computations we would like to point out that  quantity has been computed already in Refs. \cite{Ghosh:2014kba}  and \cite{Armstrong:2023phb}. We became aware of the existence of Refs. \cite{Ghosh:2014kba}  and \cite{Armstrong:2023phb} just after having obtained our result below. The reader may see in Appendix B that our result is obtained by direct calculation of the tensor algebra and integrals which occur on the right hand side of (\ref{WGRsz1z2}) and, thus, it is similar to the way of computing ${\cal W}^{(GR)}_s$ in ref. \cite{Ghosh:2014kba}. We have done however a somewhat more direct computation and we have expressed our intermediate results in terms of the Mandelstam variables and
the moduli of the vector momenta getting a compact final result. As discussed below, we have checked that our result agrees with the  one in ref. \cite{Ghosh:2014kba} when
the vector momenta $\vec{k}_a$, $a=1,2,3,4$ are symbols, ${\it i.e.}$, unknown variables. We believe this was worth-doing, for  agreement between the results in \cite{Ghosh:2014kba}  and \cite{Armstrong:2023phb} had only been checked numerically. Let us display our result:
\begin{equation}
{\cal W}^{(GR)}_s=(2\pi)^3\delta\big(\sum_{a=1}^4\vec{k}_a\big)\kappa^2 L^2\cfrac{N[k_1,k_3,k_{12},k_{34};s,t]}{D[k_1,k_3,k_{12},k_{34};s]},
\label{WGRND}
\end{equation}
where
\begin{equation}
D[k_1,k_3,k_{12},k_{34};s]=32\,\, k_{12}^3 k_{34}^3\, (k_{12}+\sqrt{s})^2\, (k_{34}+\sqrt{s})^2\,(k_{12}+k_{34})^3 s^2
\label{Dterm}
\end{equation}
and
\begin{equation}
\begin{array}{l}
{N[k_1,k_3,k_{12},k_{34};s,t]=}\\[4pt]
{(k_{12}+\sqrt{s})^2 (k_{34}+\sqrt{s})^2 (-8
   k_1^2 k_2^2 k_3^2 (k_{12}^2+3 k_{34}
   k_{12}+k_{34}^2) k_4^2 s^3+24 k_1^2 k_{12}^2
   k_2^2 k_3^2 k_{34}^2 k_4^2 s^2+2 k_{12}^2 k_3^2
   k_{34}^2\times}\\[4pt]
 {
    (k_1^2+(4 k_2+k_{34}) k_1+k_2
   (k_2+k_{34})) k_4^2 (k_1^2+k_2^2-s)
   s^2+2 k_1^2 k_{12}^2 k_2^2 k_{34}^2 (k_3^2+4
   k_4 k_3+k_4^2+k_{12} k_{34})\times}\\[4pt]
 {
   (k_3^2+k_4^2-s) s^2+2 k_3^2
   ((k_1^2+4 k_2 k_1+k_2^2) k_{12}^2+3
   (k_1^2+4 k_2 k_1+k_2^2) k_{34}
   k_{12}+2 (k_1^2+3 k_2 k_1+k_2^2)\times}\\[4pt]
 {
   k_{34}^2) k_4^2
   ((k_1^2-k_2^2)^2-(k_1^2+k_2^2)
   s) s^2+2 k_1^2 k_2^2 (2 (k_3^2+3 k_4
   k_3+k_4^2) k_{12}^2+3 k_{34} (k_3^2+4
   k_4 k_3+k_4^2) k_{12}+}\\[4pt]
   {
   k_{34}^2(k_3^2+4 k_4 k_3+k_4^2))
   ((k_3^2-k_4^2)^2-(k_3^2+k_4^2)
   s) s^2-((k_3^2+3 k_4 k_3+k_4^2)
   k_1^4+(6 k_2 (k_3^2+3 k_4
   k_3+}\\[4pt]
   {
   k_4^2)+k_{34} (2 k_3^2+7 k_4
   k_3+2 k_4^2)) k_1^3+(10 (k_3^2+3
   k_4 k_3+k_4^2) k_2^2+3 k_{34} (3
   k_3^2+11 k_4 k_3+3 k_4^2) k_2+}\\[4pt]
   {
   k_{34}^2(k_3^2+4 k_4 k_3+k_4^2)) k_1^2+3
   k_2 (2 k_2+k_{34}) ((k_2+k_{34}) k_3^2+(3
   k_2+4 k_{34}) k_4 k_3+(k_2+k_{34})
   k_4^2) k_1+}\\[4pt]
   {
   k_2^2 (k_2+k_{34})
   ((k_2+k_{34}) k_3^2+(3 k_2+4 k_{34}) k_4
   k_3+(k_2+k_{34}) k_4^2))
   ((k_1^2-k_2^2)^2-(k_1^2+k_2^2)
   s)\times}\\[4pt]
   {
   ((k_3^2-k_4^2)^2-(k_3^2+k_4^2)
   s) s-k_{12}^2 k_{34}^2 (k_{34} k_1^3+(4 k_2
   k_{34}+(2 k_3+k_4) (k_3+2 k_4)) k_1^2+(4
   k_2+k_{34})\times}\\[4pt]
 {
    (k_3^2+3 k_4
   k_3+k_4^2+k_2 k_{34}) k_1+k_2
   (k_2+k_{34}) (k_3^2+3 k_4
   k_3+k_4^2+k_2 k_{34}))
   (-((k_3^4-2 (k_4^2+s)
   k_3^2+}\\[4pt]
   {k_4^4+s^2+6 k_4^2 s) k_1^4)+(2
   (k_{34}^2+s) ((k_3-k_4)^2+s) k_2^2+s
   (2 k_3^4+(4 k_4^2-3 s-8 t) k_3^2-6
   k_4^4+s^2+}\\[4pt]
   {
   k_4^2 (5 s+8 t))) k_1^2-k_2^4
   (k_3^4-2 (k_4^2-3 s)
   k_3^2+(k_4^2-s)^2)+s^2((k_3^2+k_4^2)
   s-(k_3^2-k_4^2)^2)+k_2^2 s\times}\\[4pt]
 {
    (-6
   k_3^4+(4 k_4^2+5 s+8 t) k_3^2+2
   k_4^4+s^2-k_4^2 (3 s+8 t))))-k_{12}^3
   k_{34}^3 ((k_{34}+\sqrt{s})^2 k_1^4+(s^{3/2}+}\\[4pt]
{
   5(k_2+k_{34}) s+2 (3 k_3^2+7 k_4 k_3+3
   k_4^2+5 k_2 k_{34}) \sqrt{s}+k_{34} (5 k_2
   k_{34}+(2 k_3+k_4) (k_3+2 k_4)))
   k_1^3+}\\[4pt]
{
   (3 (k_2+k_{34}) s^{3/2}+2 (4 k_2^2+9
   k_{34} k_2+4 k_3^2+4 k_4^2+9 k_3 k_4) s+2
   (3 k_4^3+11 (k_2+k_3) k_4^2+(8 k_2^2+}\\[4pt]
{
   27k_3 k_2+11 k_3^2) k_4+k_3
   (k_2+k_3) (8 k_2+3 k_3)) \sqrt{s}+k_{34}
   (8 k_{34} k_2^2+(7 k_3^2+19 k_4 k_3+7
   k_4^2) k_2+}\\[4pt]
{
   k_{34} (k_3^2+3 k_4
   k_3+k_4^2))) k_1^2+(3 s^{3/2}
   (k_2+k_{34})^2+2 (5 k_2+k_{34}) (k_3^2+3 k_4
   k_3+k_4^2+k_2 k_{34}) \sqrt{s}\times}\\[4pt]
{
   (k_2+k_{34})+k_2 k_{34} (5 k_2+2 k_{34})
   (k_3^2+3 k_4 k_3+k_4^2+k_2
   k_{34})+(5 k_2^3+18 k_{34} k_2^2+6 (3
   k_3^2+7 k_4 k_3+}\\[4pt]
{
   3 k_4^2) k_2+k_{34}
   (5 k_3^2+13 k_4 k_3+5 k_4^2)) s)
   k_1+(k_2+k_{34}) (k_2+\sqrt{s}) (s
   (k_2+k_{34})^2+k_2 k_{34} (k_3^2+}\\[4pt]
{
   3 k_4 k_3+k_4^2+k_2 k_{34})+(2 k_2+k_{34})
   (k_3^2+3 k_4 k_3+k_4^2+k_2 k_{34})
   \sqrt{s})) ((k_3^4-2 (k_4^2+s)
   k_3^2+k_4^4+}\\[4pt]
{
   s^2+6 k_4^2 s) k_1^4-2
   (((k_3^2-k_4^2)^2-3 s^2+2
   (k_3^2+k_4^2) s) k_2^2+s (k_3^4+2
   (k_4^2-s-2 t) k_3^2-3 k_4^4+}\\[4pt]
{
   2 k_4^2 (s+2 t)+s,
   (s+4 t))) k_1^2+k_2^4 (k_3^4-2
   (k_4^2-3 s) k_3^2+(k_4^2-s)^2)-2
   k_2^2 s (-3 k_3^4+2 (k_4^2+}\\[4pt]
{
   s+2 t)k_3^2+(k_4^2-s) (k_4^2-s-4 t))+s^2
   (k_3^4+(6 k_4^2-2 (s+4 t))
   k_3^2+k_4^4+s^2+8 t^2+8 s t-}\\[4pt]
{
   2 k_4^2 (s+4 t))).
}
\label{Nterm}
\end{array}
\end{equation}
\newpage

Let $\tilde{W}^S$ and $\tilde{R}^S$ be given in (4.25) and (4.27) of reference \cite{Ghosh:2014kba}. We have verified by using Mathematica \cite{mathematica} that
\begin{equation}
\frac{1}{2}\tilde{W}^S+\tilde{ R}^S-\big(-\frac{N}{D}\big)=0,
\label{concuerda}
\end{equation}
whatever the values of $\vec{k}_1$, $\vec{k}_2$, $\vec{k}_3$ and $\vec{k}_4$, ${\it i.e.}$, when the previous vector momenta are treated as unknown variables; of course,
$\sum_{a=1}^4\vec{k}_a=0$. Note that the minus sign in front of $-N/D$ is due to the fact that our bulk-to-bulk propagator graviton is minus the bulk-to-bulk graviton propagator of ref.  \cite{Ghosh:2014kba}.

\section{Conclusions}

In this paper we have obtained by means of an explicit computation  the value, expressed in terms of the Mandelstam variables, of the momentum space $s$-channel Witten diagram corresponding to the exchange of a graviton between scalars on Euclidean Anti-de Sitter for Weyl invariant unimodular gravity. We have done this in two steps. First, we have shown by carrying out lengthy computations that the value of the Witten diagram in question is the same as in General Relativity. Then, we have computed anew the value of the
Witten diagram have just mentioned in General Relativity and obtained a rather compact result. We have verified by using Mathematica that our result agrees with the one in
ref. \cite{Ghosh:2014kba} --see (\ref{concuerda})-- for arbitrary values of the external momenta, ${\it i.e.}$, when the external momenta are symbolic unknown variables. The agreement  we have shown is far from being trivial since, in unimodular gravity, there is, on the one hand, no quadratic term on the graviton field which involves the cosmological constant and, on the other  hand, the graviton field only couples to the traceless part of the energy-momentum tensor. From a purely technical point of view notice that the axial-gauge bulk-to-bulk graviton propagator   of our unimodular theory is quite different from  its axial-gauge counterpart in General Relativity --see (\ref{BtBprop}), (\ref{coefBtBpropGR}) and (\ref{reprop}). Notice that extra poles occur in  the bulk-to-bulk propagator of the graviton of the unimodular theory considered here.

It is plain that results obtained in the previous sections can be transferred  to the  corresponding $t$-channel and $u$-channel Witten diagrams just making the substitutions $\vec{k}_2\leftrightarrow\vec{k}_{3}$ and $\vec{k}_2\leftrightarrow\vec{k}_{4}$, respectively.

The main conclusion of this paper is that Weyl invariant unimodular theory is a perfectly sensible theory at the quantum level when the cosmological constant is not zero.
Our result points in the direction that, at tree level, there will be no difference between Weyl invariant unimodular gravity and General Relativity for an anti-de Sitter or a de Sitter background: obviously, the cosmological implications of our results are the same as those from General Relativity. And yet, much further work is needed to understand the quantum properties of Weyl invariant unimodular gravity.

\section{Acknowledgements}

We should like to thank E. \'Alvarez and E. Velasco-Aja for illuminating discussions. C.P.M.'s the research work has been financially supported in part by the Spanish Ministerio de Ciencia, Innovación y Universidades under grant  PID2023-149834NB-I00.

\newpage
\appendix
\section{The integrals}
We display below  the integrals  which occur in (\ref{C1coeff}) and (\ref{C234coeff}). These integrals will be distributed in four sets and, below, we shall use the following  notation: $k_{12}=k_1+k_2$ and $k_{34}=k_3+k_4$, where $k_a$, $a=1,2,3,4$, is the modulus of $\vec{k}_a$. As usual $s=(\vec{k}_1+\vec{k}_2)^2$ is the Mandelstam $s$-variable.
Bear also in mind the $G^{(b)}(z,\vec{k})$ is the boundary-to-bulk propagator in (\ref{btBGreen}).

\subsection*{Set I}
Let
\begin{equation}
f_{1}(\omega;p,q)=\int_{0}^{\infty}\!\!\!dz\, \partial_z G^{(b)}(z,\vec{p})\,\partial_z G^{(b)}(z,\vec{q})\, z^{-1/2} J_{3/2}[\omega z].
\label{f1}
\end{equation}
Then, we define
\begin{equation*}
\begin{array}{l}
{
\text{I}_{11}=-\int_0^\infty\!\!d\omega\,\cfrac{\omega}{s^2(s+\omega^2)}\,f_{1}(\omega;k_1,k_2)\,f_{1}(\omega;k_3,k_4),}\\[2pt]
{
\text{I}_{12}=-\lim_{\epsilon\rightarrow 0}\int_0^\infty\!\!d\omega\,\cfrac{\omega}{s^2(s+\omega^2)}\;\cfrac{2(s^2-3\omega^4)}{((\sqrt{s}-\omega-i\epsilon)(\sqrt{s}+\omega-i\epsilon))^2}\,f_{1}(\omega;k_1,k_2)\,f_{1}(\omega;k_3,k_4),\quad
}\\[2pt]
{
\text{I}_{13}=\lim_{\epsilon\rightarrow 0}\int_0^\infty\!\!d\omega\,\cfrac{\omega}{s^2(s+\omega^2)}\;
\cfrac{4 s\omega^2}{((\sqrt{s}-\omega-i\epsilon)(\sqrt{s}+\omega-i\epsilon))^2}\,f_{1}(\omega;k_1,k_2)\,f_{1}(\omega;k_3,k_4),
}\\[2pt]
{
\text{I}_{14}=-\int_0^\infty\!\!d\omega\,\cfrac{\omega}{s^2(s+\omega^2)}\;\cfrac{s}{\omega^2}\,f_{1}(\omega;k_1,k_2)\,f_{1}(\omega;k_3,k_4),
}\\[2pt]
{
\text{I}_{15}=\lim_{\epsilon\rightarrow 0}\int_0^\infty\!\!d\omega\,\cfrac{\omega}{s^2(s+\omega^2)}\;
\cfrac{4 s^3}{\omega^2((\sqrt{s}-\omega-i\epsilon)(\sqrt{s}+\omega-i\epsilon))^2}\,f_{1}(\omega;k_1,k_2)\,f_{1}(\omega;k_3,k_4),
}\\[2pt]
{
\text{I}_{16}=\lim_{\epsilon\rightarrow 0}\int_0^\infty\!\!d\omega\,\cfrac{\omega}{s^2(s+\omega^2)}\;
\cfrac{-3 s^4+2s^3\omega^2+5s^2\omega^4}{((\sqrt{s}-\omega-i\epsilon)(\sqrt{s}+\omega-i\epsilon))^2}\,f_{1}(\omega;k_1,k_2)\,f_{1}(\omega;k_3,k_4),
}\\[2pt]
{
\text{I}_{17}=\lim_{\epsilon\rightarrow 0}\int_0^\infty\!\!d\omega\,
\cfrac{\omega(-s+3\omega^2)}{\omega^2((\sqrt{s}-\omega-i\epsilon)(\sqrt{s}+\omega-i\epsilon))^2}\,f_{1}(\omega;k_1,k_2)\,f_{1}(\omega;k_3,k_4),
}\\[2pt]
{
\text{I}_{18}=\lim_{\epsilon\rightarrow 0}\int_0^\infty\!\!d\omega\,
\cfrac{\omega(s+\omega^2)}{\omega^4((\sqrt{s}-\omega-i\epsilon)(\sqrt{s}+\omega-i\epsilon))^2}\,f_{1}(\omega;k_1,k_2)\,f_{1}(\omega;k_3,k_4).
}\\[2pt]
\end{array}
\end{equation*}
The computation of the previous integrals yields the following results:
\begin{equation*}
\begin{array}{l}
{\text{I}_{11}=- \cfrac{2 k_1^2 k_2^2 k_3^2 k_4^2 \bigl(k_{12} k_{34} + 2 (k_{12} +k_{34}) \sqrt{s} + s\bigr)}{(k_{12} + k_{34})^3 (k_{12} + \sqrt{s})^2 (k_{34} + \sqrt{s})^2 s^2}}
\end{array}
\end{equation*}
\begin{equation*}
\begin{array}{l}
{\text{I}_{12}=\cfrac{4 k_1^2 k_2^2 k_3^2 k_4^2 A_{12}[k_1,k_2,k_3,k_4] }{(k_{12} + k_{34})^3 (k_{12}+ i \sqrt{s})^3 (k_{34} + i \sqrt{s})^3 (k_{12}+ \sqrt{s})^2 (k_{34} + \sqrt{s})^2 s^2}}\\[4pt]
\end{array}
\end{equation*}
with
\begin{equation*}
\begin{array}{l}
{A_{12}=3 k_{12}^4 k_{34}^4 + (6+9i) k_{12}^3 k_{34}^3 (k_{12} + k_{34}) \sqrt{s} + (-3+18i) k_{12}^2 k_{34}^2 (k_{12} + k_{34})^2 s +}\\[4pt]
{+ i k_{12} k_{34} (k_{12} + k_{34}) \bigl((7+11i) k_{12}^2 + (23+28i) k_{12} k_{34} + (7+11i) k_{34}^2\bigr) \sqrt{s^3} - \bigl((4+i) k_{12}^4+}\\[4pt]
{ + (28+4i) k_{12}^3 k_{34} + (52+6i) k_{12}^2 k_{34}^2 + (28+4i)k_{12} k_{34}^3 +(4+i) k_{34}^4\bigr) s^2 - (k_{12} + k_{34})\times }\\[4pt]
{\times\bigl((5+5i) k_{12}^2 + (12+13i) k_{12} k_{34} + (5+5i) k_{34}^2\bigr) \sqrt{s^5} + (-1-6i) (k_{12} + k_{34})^2 s^3 +}\\[4pt]
{ +(2-3i) (k_{12} + k_{34}) \sqrt{s^7} + s^4}
\end{array}
\end{equation*}
\begin{equation*}
\begin{array}{l}
{\text{I}_{13}=\cfrac{4 k_1^2 k_2^2 k_3^2 k_4^2 A_{13}[k_1,k_2,k_3,k_4]}{(k_{12} + k_{34})^3 (k_{12} + i \sqrt{s})^3 (k_{34} + i \sqrt{s})^3 s \bigl(k_{12} k_{34}+ (k_{12} + k_{34}) \sqrt{s} + s\bigr)^2}}
\end{array}
\end{equation*}
with
\begin{equation*}
\begin{array}{l}
{A_{13}=2 k_{12}^2 k_{34}^2 (k_{12}^2 + 3 k_{12}k_{34} + k_{34}^2) +k_{12} k_{34} (k_{12} + k_{34}) \bigl((3+2i) k_{12}^2 +10 (1+i) k_{12} k_{34} +}\\[4pt]
{+ (3+2i) k_{34}^2\bigr) \sqrt{s} + i (k_{12} +k_{34})^2 (k_{12}^2 + 14 k_{12} k_{34} + k_{34}^2) s + i (k_{12}+ k_{34}) \bigl((2+3i) k_{12}^2+}\\[4pt]
{ + 10(1+i) k_{12}k_{34} + (2+3i) k_{34}^2\bigr) \sqrt{s^3} - 2 (k_{12}^2 + 3 k_{12} k_{34} + k_{34}^2) s^2}
\end{array}
\end{equation*}
\begin{equation*}
\begin{array}{l}
{\text{I}_{14}=- \cfrac{2 k_1^2 k_2^2 k_3^2 k_4^2 \bigl(k_{12}^2 + 3 k_{12} k_{34} + k_{34}^2 + 2 (k_{12} + k_{34}) \sqrt{s} + s\bigr)}{k_{12}k_{34} (k_{12} + k_{34})^3 s \bigl(k_{12} k_{34} + (k_{12}+ k_{34}) \sqrt{s}  + s\bigr)^2}}
\end{array}
\end{equation*}
\begin{equation*}
\begin{array}{l}
{\text{I}_{15}=\cfrac{4 k_1^2 k_2^2 k_3^2 k_4^2 A_{15}[k_1,k_2,k_3,k_4]}{k_{12} k_{34} (k_{12} + k_{34})^3 (k_{12} + i \sqrt{s})^3 (k_{34} + i \sqrt{s})^3 \sqrt{s}\bigl(k_{12} k_{34} + (k_{12} + k_{34}) \sqrt{s} + s\bigr)^2}}\\[4pt]
\end{array}
\end{equation*}
with
\begin{equation*}
\begin{array}{l}
{A_{15}=- k_{12}^2 k_{34}^2 (k_{12} + k_{34})^3 + (-2-3i)k_{12}k_{34} (k_{12} + k_{34})^4 \sqrt{s} - i (k_{12} + k_{34})^3 \bigl(2 k_{12}^2+}\\[4pt]
{ + (12+5i) k_{12}k_{34} + 2 k_{34}^2\bigr) s - 2i \bigl((2+3i) k_{12}^4 + (8+18i) k_{12}^3 k_{34} + (12+31i) k_{12}^2 k_{34}^2 + }\\[4pt]
{+(8+18i) k_{12} k_{34}^3 + (2+3i) k_{34}^4\bigr) \sqrt{s^3}+2 (k_{12} + k_{34}) \bigl((6+2i) k_{12}^2 + (14+7i) k_{12} k_{34} + }\\[4pt]
{+(6+2i) k_{34}^2\bigr) s^2 + (4+12i) (k_{12} + k_{34})^2 \sqrt{s^5} + (-4+6i) (k_{12} + k_{34}) s^3 - 2 \sqrt{s^7}}
\end{array}
\end{equation*}
\begin{equation*}
\begin{array}{l}
{\text{I}_{16}=\cfrac{2 k_1^2 k_2^2 k_3^2 k_4^2 A_{16}[k_1,k_2,k_3,k_4]}{(k_{12}  + k_{34})^3 (k_{12}  + i \sqrt{s})^3 (k_{34} + i \sqrt{s})^3}}\\[4pt]
\end{array}\end{equation*}
with
\begin{equation*}
\begin{array}{l}
{A_{16}=5 k_{12} ^2 k_{34}^2 + 15i k_{12}  k_{34}(k_{12}  + k_{34}) \sqrt{s} - 8 (k_{12} ^2 + 3 k_{12}  k_{34} + k_{34}^2) s -}\\[4pt]
{- 9i (k_{12}  + k_{34}) \sqrt{s^3} + 3 s^2}
\end{array}
\end{equation*}
\begin{equation*}
\begin{array}{l}
{\text{I}_{17}=\cfrac{2 k_1^2 k_2^2 k_3^2 k_4^2 A_{17}[k_1,k_2,k_3,k_4]}{k_{12} k_{34} (k_{12} + k_{34})^3 (k_{12} + i \sqrt{s})^3 (k_{34} + i \sqrt{s})^3}}\\[4pt]
\end{array}\end{equation*}
with
\begin{equation*}
\begin{array}{l}
{A_{17}=3 k_{12} k_{34} (k_{12}^2 + 3k_{12} k_{34} + k_{34}^2) + i (k_{12} + k_{34}) (k_{12}^2 + 11 k_{12} k_{34} + k_{34}^2) \sqrt{s} -}\\[4pt]
{-  (3 k_{12} +k_{34}) (k_{12} + 3 k_{34}) s - 3i (k_{12} + k_{34}) \sqrt{s^3} + s^2}
\end{array}
\end{equation*}
\begin{equation*}
\begin{array}{l}
{\text{I}_{18}=\cfrac{2 k_1^2 k_2^2 k_3^2 k_4^2 A_{18}[k_1,k_2,k_3,k_4]}{(k_{12}^2 -  k_{34}^2)^3 }}
\end{array}
\end{equation*}
with
\begin{equation*}
\begin{array}{l}
{A_{18}=- \cfrac{2 (k_{12}^2 -  k_{34}^2) }{k_{12}^2 (k_{12} + i \sqrt{s})^3} -  \cfrac{5 k_{12}^2 -  k_{34}^2}{k_{12}^3 (k_{12} + i \sqrt{s})^2} -  \cfrac{2 (k_{12}^2 -  k_{34}^2) }{k_{34}^2 (k_{34} + i \sqrt{s})^3} + \cfrac{k_{12}^2 - 5 k_{34}^2}{k_{34}^3 (-i k_{34} + \sqrt{s})^2}}\\[4pt]
\end{array}
\end{equation*}
\subsection*{Set II}
We introduce next
\begin{equation}
f_{2}(\omega;p,q)=\int_{0}^{\infty}\!\!\!dz\,G^{(b)}(z,\vec{p})\, G^{(b)}(z,\vec{q})\, z^{-1/2} J_{3/2}[\omega z]
\label{f2}
\end{equation}
and
\begin{equation*}
\begin{array}{l}
{
\text{I}_{21}=-\int_0^\infty\!\!d\omega\,\cfrac{\omega}{s^2(s+\omega^2)}\,f_{1}(\omega;k_1,k_2)\,f_{2}(\omega;k_3,k_4),}\\[2pt]
{
\text{I}_{22}=-\lim_{\epsilon\rightarrow 0}\int_0^\infty\!\!d\omega\,\cfrac{\omega}{s^2(s+\omega^2)}\;\cfrac{2(s^2-3\omega^4)}{((\sqrt{s}-\omega-i\epsilon)(\sqrt{s}+\omega-i\epsilon))^2}\,f_{1}(\omega;k_1,k_2)\,f_{2}(\omega;k_3,k_4),\quad
}\\[2pt]
{
\text{I}_{23}=\lim_{\epsilon\rightarrow 0}\int_0^\infty\!\!d\omega\,\cfrac{\omega}{s^2(s+\omega^2)}\;
\cfrac{4 s\omega^2}{((\sqrt{s}-\omega-i\epsilon)(\sqrt{s}+\omega-i\epsilon))^2}\,f_{1}(\omega;k_1,k_2)\,f_{2}(\omega;k_3,k_4),
}\\[2pt]
{
\text{I}_{24}=-\int_0^\infty\!\!d\omega\,\cfrac{\omega}{s^2(s+\omega^2)}\;\cfrac{s}{\omega^2}\,f_{1}(\omega;k_1,k_2)\,f_{2}(\omega;k_3,k_4),
}\\[2pt]
{
\text{I}_{25}=\lim_{\epsilon\rightarrow 0}\int_0^\infty\!\!d\omega\,\cfrac{\omega}{s^2(s+\omega^2)}\;
\cfrac{4 s^3}{\omega^2((\sqrt{s}-\omega-i\epsilon)(\sqrt{s}+\omega-i\epsilon))^2}\,f_{1}(\omega;k_1,k_2)\,f_{2}(\omega;k_3,k_4),
}\\[2pt]
{
\text{I}_{26}=\lim_{\epsilon\rightarrow 0}\int_0^\infty\!\!d\omega\,\cfrac{\omega}{s^2(s+\omega^2)}\;
\cfrac{-3 s^4+2s^3\omega^2+5s^2\omega^4}{((\sqrt{s}-\omega-i\epsilon)(\sqrt{s}+\omega-i\epsilon))^2}\,f_{1}(\omega;k_1,k_2)\,f_{2}(\omega;k_3,k_4),
}\\[2pt]
{
\text{I}_{27}=\lim_{\epsilon\rightarrow 0}\int_0^\infty\!\!d\omega\,
\cfrac{\omega(-s+3\omega^2)}{\omega^2((\sqrt{s}-\omega-i\epsilon)(\sqrt{s}+\omega-i\epsilon))^2}\,f_{1}(\omega;k_1,k_2)\,f_{2}(\omega;k_3,k_4),
}\\[2pt]
{
\text{I}_{28}=\lim_{\epsilon\rightarrow 0}\int_0^\infty\!\!d\omega\,
\cfrac{\omega(s+\omega^2)}{\omega^4((\sqrt{s}-\omega-i\epsilon)(\sqrt{s}+\omega-i\epsilon))^2}\,f_{1}(\omega;k_1,k_2)\,f_{2}(\omega;k_3,k_4),
}\\[2pt]
\end{array}
\end{equation*}
where $f_{1}(\omega;k_3,k_4)$ has been defined in (\ref{f1}).

We have worked out the values of the previous integrals, which read
\begin{equation*}
\begin{array}{l}
{\text{I}_{21}=\cfrac{k_1^2 k_2^2 A_{21}[k_1,k_2,k_3,k_4]}{(k_{12}+k_{34})^3 (k_{12} + \sqrt{s})^2 (k_{34} + \sqrt{s})^2 s^2}}
\end{array}
\end{equation*}
with
\begin{equation*}
\begin{array}{l}
{A_{21}=- k_{12} k_{34} (k_3^2 + k_{12} k_{34} + 4 k_3 k_4 + k_4^2) - 2 (k_{12} + k_{34}) (k_3^2 + k_{12} k_{34} + 4 k_3 k_4 + k_4^2) \sqrt{s} -  }\\[4pt]
{-\bigl(k_{12}^2 + 3 k_{12} k_{34}+ 2 (k_3^2 + 3 k_3 k_4 + k_4^2)\bigr) s}
\end{array}
\end{equation*}
\begin{equation*}
\begin{array}{l}
{\text{I}_{22}=\cfrac{2 k_1^2 k_2^2 A_{22}[k_1,k_2,k_3,k_4]}{(k_{12} +  k_{34})^3 (k_{12} + i \sqrt{s})^3 ( k_{34} + i \sqrt{s})^3 (k_{12}+ \sqrt{s})^2 ( k_{34} + \sqrt{s})^2 s^2}}\\[4pt]
\end{array}
\end{equation*}
with
\begin{equation*}
\begin{array}{l}
{A_{22}=3 k_{12}^4 k_{34}^4 \bigl(k_3^2 + 4 k_3 k_4 + k_4^2 + k_{12}k_{34}\bigr) + (6+9i) k_{12}^3 k_{34}^3 (k_{12} + k_{34}) (k_3^2 + k_{12} k_{34}+ 4 k_3 k_4  +}\\[4pt]
{+ k_4^2) \sqrt{s}+ 3i k_{12}^2 k_{34}^2 (k_{12} + k_{34})^2 \bigl((6+2i) k_{12} k_{34} + (6+i) (k_3^2 + 4 k_3 k_4 + k_4^2)\bigr) s + i k_{12} k_{34} (k_{12}+}\\[4pt]
{  + k_{34})\Bigl[(6+18i) k_{12}^3k_{34}+ (7+11i) k_{34}^2 (k_3^2 + 4 k_3 k_4 + k_4^2) + k_{12}^2 \bigl((28+53i) k_3^2 + (70+}\\[4pt]
{+128i) k_3 k_4 + (28+53i) k_4^2\bigr) + k_{12} k_{34} \bigl((29+46i) k_3^2 + (104+148i) k_3 k_4 + (29+46i) k_4^2\bigr)\Bigr] \sqrt{s^3} - }\\[4pt]
{-\Bigl[(9+6i) k_1^5 k_3 + k_1^4 k_3 \bigl((45+30i) k_2 + (59+25i) k_3\bigr) + 2 k_1^3 k_3 \bigl((45+30i) k_2^2+(118+}\\[4pt]
{+50i) k_2 k_3 + (62+20i) k_3^2\bigr)  + k_1^2 k_3 \bigl((90+60i) k_2^3 + (354+150i) k_2^2 k_3 + (372+120i) k_2 k_3^2 +}\\[4pt]
{+ (107+30i) k_3^3\bigr)  + k_3k_{23}  \bigl((9+6i) k_2^4+ (50+19i) k_2^3 k_3 + (74+21i) k_2^2 k_3^2 + (33+9i) k_2 k_3^3  + }\\[4pt]
{ + (4+i) k_3^4\bigr)+k_1 k_3 \bigl((45+30i) k_2^4+ (236+100i) k_2^3 k_3 +(372+120i) k_2^2 k_3^2 + (214 +}\\[4pt]
{+60i) k_2 k_3^3+ (37+10i) k_3^4\bigr)+ (9+6i) k_{12}^5 k_4  + k_3 \bigl((126+52i) k_{12}^4 + (428+128i) k_{12}^3 k_3 +}\\[4pt]
{+(532+132i) k_{12}^2 k_3^2 + (241+58i) k_{12} k_3^3 + (32+8i) k_3^4\bigr) k_4 + \bigl((59+25i) k_{12}^4 + (428+ }\\[4pt]
{+128i) k_{12}^3 k_3 + (850+204i)k_{12}^2 k_3^2 + (538+124i) k_{12} k_3^3 + (92+23i) k_3^4\bigr) k_4^2 +2 \bigl((62+ }\\[4pt]
{ +20i)k_{12}^3 + (266+66i) k_{12}^2 k_3 + (269+62i) k_{12} k_3^2 + (64+16i) k_3^3\bigr) k_4^3+\bigl((107+30i) k_{12}^2 + }\\[4pt]
{+ (241+58i) k_{12} k_3 + (92+23i) k_3^2\bigr) k_4^4 + \bigl((37+10i) k_{12}  + (32+8i) k_3\bigr) k_4^5+ \bigl((37+10i) k_{12} +}\\[4pt]
{+(32+8i) k_3\bigr) k_4^5+ (4+i) k_4^6\Bigr] s^2 -  (k_{12} + k_{34}) \Bigl[3i k_1^4 + 3i k_2^4 + (9+25i) k_2^3 k_3 + (25+}\\[4pt]
{+52i) k_2^2 k_3^2 + (21+38i) k_2 k_3^3 + (5+8i) k_3^4 + k_1^3 \bigl(12i k_2 + (9+25i) k_3\bigr) + k_1^2 \bigl(18i k_2^2 +}\\[4pt]
{+ (27+75i) k_2 k_3  + (25+52i) k_3^2\bigr)+ k_1 \bigl(12i k_2^3 + (27+75i) k_2^2 k_3 + (50+104i) k_2 k_3^2 + }\\[4pt]
{ + (21+38i) k_3^3\bigr) + (9+25i) k_{12}^3 k_4+k_3 \bigl((60+114i) k_{12}^2 + (87+140i) k_{12} k_3 + (30+ }\\[4pt]
{ +42i) k_3^2\bigr) k_4 + \bigl((25+52i) k_{12}^2 + (87+140i) k_{12} k_3 + (50+68i) k_3^2\bigr) k_4^2 + \bigl((21+38i) k_{12}+  }\\[4pt]
{+(30+42i) k_3\bigr) k_4^3 +(5+8i) k_4^4\Bigr] \sqrt{s^5} -i (k_{12} + k_{34})^2 \bigl((5+3i) k_{12}^2 + (11+2i) k_3^2  + }\\[4pt]
{+(16+6i)k_{12}k_{34}+(34+2i) k_3 k_4 - (11+2i) k_4^2\bigr) s^3 - i (k_{12} +k_{34}) \bigl((1+3i) k_{12}^2 + }\\[4pt]
{+(4+5i) k_3^2  + (5+8i) k_{12}k_{34}+ (14+14i) k_3 k_4 +(4+5i) k_4^2\bigr) \sqrt{s^7} + }\\[4pt]
{+\bigl(k_{12}^2 +3 k_{12}k_{34} + 2 (k_3^2 + 3 k_3 k_4 + k_4^2)\bigr) s^4}
\end{array}
\end{equation*}
\begin{equation*}
\begin{array}{l}
{\text{I}_{23}=\cfrac{k_1^2 k_2^2 A_{23}[k_1,k_2,k_3,k_4]}{(k_{12}^2 -  k_{34}^2)^3  s}}
\end{array}
\end{equation*}
with
\begin{equation*}
\begin{array}{l}
{A_{23}=- \cfrac{2 (k_{12}^2 -  k_{34}^2)  (k_{12}^2 -  k_3^2 - 4 k_3 k_4 -  k_4^2)}{(k_{12} + i \sqrt{s})^3} - \cfrac{1}{k_{12} (k_{12} + i \sqrt{s})^2} \Big[3 k_1^4 + 12 k_1^3 k_2 + 3 k_2^4 + k_{34}^2 (k_3^2  + }\\[4pt]
{+ 4 k_3 k_4+k_4^2) - 2 k_2^2 (2 k_3^2 + 9 k_3 k_4 + 2 k_4^2) - 2 k_1^2 (-9 k_2^2 + 2 k_3^2 + 9 k_3 k_4 + 2 k_4^2) - 4 k_1 k_2 (-3 k_2^2  + }\\[4pt]
{+ 2 k_3^2+9 k_3 k_4 + 2 k_4^2)\Big]+ \cfrac{2 (k_{12}^2 -  k_3^2 - 6 k_3 k_4 -  k_4^2)}{k_{34} + \sqrt{s}}+ \cfrac{4 (- k_{12}^2 + k_3^2 + 6 k_3 k_4 + k_4^2)}{k_{12} + i \sqrt{s}}  + }\\[4pt]
{ + \cfrac{2 \bigl(k_{12}^2 (k_3^2 + 3 k_3 k_4 + k_4^2) -  k_{34}^2 (k_3^2 + 7 k_3 k_4 + k_4^2)\bigr)}{k_{34} (k_{34} + i\sqrt{s})^2} -  \cfrac{(k_{12}^2 -  k_{34}^2)  (k_{12}^2 -  k_3^2 - 4 k_3 k_4 -  k_4^2)}{k_{12} (k_{12} + \sqrt{s})^2} +}\\[4pt]
{+ \cfrac{2 (- k_{12}^2 + k_3^2 + 6 k_3 k_4 + k_4^2)}{k_{12} + \sqrt{s}} + \cfrac{2 k_3 (k_{12}^2 -  k_{34}^2) k_4}{k_{34} (k_{34} + \sqrt{s})^2}+ \cfrac{4 k_3 (k_{12}^2 -  k_{34}^2)  k_4}{(k_{34} + i \sqrt{s})^3}+\cfrac{4 (k_{12}^2 -  k_3^2 - 6 k_3 k_4 -  k_4^2)}{k_{34} + i \sqrt{s}} }
\end{array}
\end{equation*}
\begin{equation*}
\begin{array}{l}
{\text{I}_{24}=- \cfrac{k_1^2 k_2^2A_{24}[k_1,k_2,k_3,k_4]}{k_{12} k_{34} (k_{12} + k_{34})^3 s \bigl(k_{12}k_{34} + (k_{12} + k_{34})\sqrt{s} + s\bigr)^2}}
\end{array}
\end{equation*}
with
\begin{equation*}
\begin{array}{l}
{A_{24}=(k_3^2 + 4 k_3 k_4 + k_4^2) (k_{34} + \sqrt{s})^2 + 2 k_1^2 (k_3^2 + 3 k_3 k_4 + k_4^2 + k_{34} \sqrt{s})+2 k_2^2 (k_3^2 + 3 k_3 k_4 + }\\[4pt]
{ + k_4^2 + k_{34}\sqrt{s}) +k_2 \bigl(3 k_{34} (k_3^2 + 4 k_3 k_4 + k_4^2) + 4 (k_3^2 + 3 k_3 k_4 + k_4^2)\sqrt{s} + k_{34} s\bigr) +k_1 \bigl(3 k_{34} (k_3^2 + }\\[4pt]
{+ 4 k_3 k_4 + k_4^2) + 4 k_2 (k_3^2 + 3 k_3 k_4 + k_4^2 + k_{34} \sqrt{s}) + 4 (k_3^2 + 3 k_3 k_4 + k_4^2) \sqrt{s} + k_{34} s\bigr)}
\end{array}
\end{equation*}
\begin{equation*}
\begin{array}{l}
{\text{I}_{25}=k_1^2 k_2^2 \Bigg[\cfrac{2 (k_{12}^2 -  k_3^2 - 4 k_3 k_4 -  k_4^2)}{(k_{12}^3 -  k_{12} k_{34}^2)^2 (k_{12} + i \sqrt{s})^3} +\cfrac{1}{(k_{12}^3 -  k_{12} k_{34}^2)^3 (k_{12} + i \sqrt{s})^2}\Big[3 k_1^4 + 12 k_1^3 k_2 + }\\[4pt]
{+ 3 k_2^4 - 4 k_2^2 k_3^2 + k_3^4 +2 k_1^2 (9 k_2^2 - 2 k_3^2) + 4 k_1 (3 k_2^3 - 2 k_2 k_3^2) + 6 k_3 (-3 k_{12}^2 + k_3^2) k_4 +2 (-2 k_{12}^2  + }\\[4pt]
{+ 5 k_3^2) k_4^2+ 6 k_3 k_4^3  +k_4^4\Big]- \cfrac{4 k_3 k_4}{(- k_{12}^2 k_{34} + k_{34}^3)^2 (k_{34} + i \sqrt{s})^3}-  \cfrac{2 (k_3^2 + 4 k_3 k_4 + k_4^2)}{k_{12}^4 k_{34}^4 \sqrt{s}}- }\\[4pt]
{ -\cfrac{k_{12}^2 -  k_3^2 - 4 k_3 k_4 -  k_4^2}{k_{12}^3 (k_{12}^2 -  k_{34}^2)^2 (k_{12} +\sqrt{s} )^2} -   \cfrac{2 \bigl(2 k_{12}^4 - 3 k_{12}^2 (k_3^2 + 4 k_3 k_4 + k_4^2) + k_{34}^2 (k_3^2 + 4 k_3 k_4 + k_4^2)\bigr)}{k_{12}^4 (k_{12}^2 -  k_{34}^2)^3 (k_{12} + \sqrt{s})} +}\\[4pt]
{+ \cfrac{2 k_3 k_4}{k_{34}^3 (k_{12}^2 - k_{34}^2)^2 (k_{34} + \sqrt{s})^2} + \cfrac{2 k_{12}^2 (k_3^2 + 4 k_3 k_4 + k_4^2) - 2 k_{34}^2 (k_3^2 + 8 k_3 k_4 + k_4^2)}{k_{34}^4 (k_{12}^2 -  k_{34}^2)^3 (k_{34} + \sqrt{s})} +}\\[4pt]
{+\cfrac{2 \bigl(k_{12}^2 (k_3^2 + 3 k_3 k_4 + k_4^2) -  k_{34}^2 (k_3^2 + 7 k_3 k_4 + k_4^2)\bigr)}{(- k_{12}^2 k_{34} + k_{34}^3)^3 (k_{34} + i \sqrt{s})^2}\Bigg]}
\end{array}
\end{equation*}
\begin{equation*}
\begin{array}{l}
{\text{I}_{26}=\cfrac{k_1^2 k_2^2 A_{26}[k_1,k_2,k_3,k_4]}{(k_{12} + k_{34})^3 (k_{12} + i \sqrt{s})^3 (k_{34} + i \sqrt{s})^3}}
\end{array}
\end{equation*}
with
\begin{equation*}
\begin{array}{l}
{A_{26}=5 k_2^2 k_{34}^2 (k_3^2 + k_2 k_{34} + 4 k_3 k_4 + k_4^2) + 5 k_1^3 (k_{34} + i \sqrt{s})^3 + 15i k_2 k_{34} (k_2 + k_{34}) (k_3^2  +}\\[4pt]
{+ k_2 k_{34}+ 4 k_3 k_4 + k_4^2) \sqrt{s}-  \bigl(15 k_2^3 k_{34} + 8 k_{34}^2 (k_3^2 + 4 k_3 k_4 + k_4^2) + 3 k_2 k_{34} (13 k_3^2 + 42 k_3 k_4 + }\\[4pt]
{+ 13 k_4^2) +2 k_2^2 (23 k_3^2 + 54 k_3 k_4 + 23 k_4^2)\bigr) s -i (k_2 + k_{34}) (5 k_2^2 + 14 k_3^2 + 19 k_2k_{34}+ 46 k_3 k_4 +}\\[4pt]
{+ 14 k_4^2) \sqrt{s^3} + 3 \bigl(k_2^2 + 3 k_2 k_{34}+ 2 (k_3^2 + 3 k_3 k_4 + k_4^2)\bigr) s^2 + k_1^2 \bigl[ k_{34}^2 (k_3^2 + 4 k_3 k_4 + k_4^2) +}\\[4pt]
{ + 15 k_2 (k_{34} + i\sqrt{s})^3+30i k_{34} (k_3^2 + 3 k_3 k_4 + k_4^2) \sqrt{s}- 2 (23 k_3^2 + 54 k_3 k_4 + 23 k_4^2) s -}\\[4pt]
{ -24i k_{34} \sqrt{s^3}+ 3 s^2\bigr] + k_1 \Bigl[15 k_2^2 (k_{34} + i \sqrt{s})^3 + 3i (k_3^2 + 4 k_3 k_4 + k_4^2 + i k_{34}\sqrt{s}) (5k_{34}^2 + }\\[4pt]
{ + 8i k_{34} \sqrt{s} - 3 s) \sqrt{s}+ 2 k_2 \bigl(5 k_{34}^2 (k_3^2 +4 k_3 k_4 + k_4^2) + 30i k_{34} (k_3^2 + 3 k_3 k_4 + k_4^2)\sqrt{s} -  }\\[4pt]
{-2 (23 k_3^2 + 54 k_3 k_4 + 23 k_4^2) s - 24i k_{34} \sqrt{s^3} + 3 s^2\bigr)\Bigr]}
\end{array}
\end{equation*}
\begin{equation*}
\begin{array}{l}
{\text{I}_{27}=\cfrac{k_1^2 k_2^2 A_{27}[k_1,k_2,k_3,k_4] }{k_{12} k_{34} (k_{12} + k_{34})^3 (k_{12} + i \sqrt{s})^3 (k_{34} + i\sqrt{s})^3}}
\end{array}
\end{equation*}
with
\begin{equation*}
\begin{array}{l}
{A_{27}=3 k_2^2 (k_3^2 + 4 k_3 k_4 + k_4^2 + i k_{34} \sqrt{s}) (3 k_{34}^2 + 4i k_{34} \sqrt{s} -  s) + i (k_3^2 + 4 k_3 k_4 + k_4^2) (k_{34} + }\\[4pt]
{+ i \sqrt{s})^3 \sqrt{s} +2( k_1^3+k_2^3) \bigl(3 k_{34} (k_3^2 + 3 k_3 k_4 + k_4^2) + i (5 k_3^2 + 11 k_3 k_4 + 5 k_4^2) \sqrt{s}-2 k_{34} s\bigr)+ }\\[4pt]
{ + 3 k_1^2 \bigl(6 k_2 k_{34}(k_3^2 + 3 k_3 k_4 + k_4^2)+ (k_3^2 + 4 k_3 k_4 + k_4^2 + i k_{34} \sqrt{s}) (3 k_{34}^2 + 4i k_{34}\sqrt{s} -  s) +}\\[4pt]
{+2i k_2 (5 k_3^2 + 11 k_3 k_4 + 5 k_4^2) \sqrt{s} - 4 k_2 k_{34} s\bigr) + k_2 \bigl[3 k_{34}^3 (k_3^2 + 4 k_3 k_4 + k_4^2) + 12i k_{34}^2 (k_3^2 +}\\[4pt]
{ + 4 k_3 k_4 + k_4^2) \sqrt{s} -2 k_{34} (7 k_3^2 + 24 k_3 k_4 + 7 k_4^2) s-  6i (k_3^2 + 3 k_3 k_4 + k_4^2) \sqrt{s^3} + k_{34} s^2\bigr]   +}\\[4pt]
{+ k_1 \Bigl[3 k_{34}^3 (k_3^2 + 4 k_3 k_4 + k_4^2) + 6 k_2 (k_3^2 + 4 k_3 k_4 + k_4^2+ i k_{34} \sqrt{s}) (3 k_{34}^2 + 4i k_{34} \sqrt{s} -  s)+}\\[4pt]
{+ 12i k_{34}^2 (k_3^2 + 4 k_3 k_4 + k_4^2) \sqrt{s} - 2 k_{34}(7 k_3^2 + 24 k_3 k_4 + 7 k_4^2) s - 6i (k_3^2 + 3 k_3 k_4 + k_4^2) \sqrt{s^3} + }\\[4pt]
{+k_{34}s^2 + 6 k_2^2 \bigl(3 k_{34} (k_3^2 + 3 k_3 k_4 + k_4^2) + i (5 k_3^2 + 11 k_3 k_4 + 5 k_4^2) \sqrt{s} - 2 k_{34} s\bigr)\Bigr]}\\[4pt]
\end{array}\end{equation*}
\begin{equation*}
\begin{array}{l}
{\text{I}_{28}=\cfrac{k_1^2 k_2^2 }{(k_{12}^2 -  k_{34}^2)^3 }\Bigg[\cfrac{2 (k_{12}^2 -  k_{34}^2)  (k_{12}^2 -  k_3^2 - 4 k_3 k_4 -  k_4^2)}{k_{12}^2 (k_{12} + i \sqrt{s})^3}+ \cfrac{1}{k_{12}^3 (k_{12} + i \sqrt{s})^2}\Big[3 k_1^4 + 12 k_1^3 k_2  +}\\[4pt]
{+3 k_2^4 - 4 k_2^2 k_3^2 + k_3^4 + 2 k_1^2 (9 k_2^2 - 2 k_3^2)+4 k_1 (3 k_2^3 - 2 k_2 k_3^2) + 6 k_3 (-3 k_{12}^2 + k_3^2) k_4 + 2 (-2 k_{12}^2 + }\\[4pt]
{+5 k_3^2) k_4^2  + 6 k_3 k_4^3 + k_4^4\Big] - \cfrac{4 k_3 (k_{12}^2 - k_{34}^2)  k_4}{k_{34}^2 (k_{34} + i \sqrt{s})^3} + \cfrac{2 \bigl(- k_{12}^2 (k_3^2 + 3 k_3 k_4 + k_4^2) + k_{34}^2 (k_3^2 + 7 k_3 k_4 + k_4^2)\bigr)}{k_{34}^3 (k_{34} + i \sqrt{s})^2}\Bigg]}
\end{array}
\end{equation*}
\subsection*{Set III}
The integrals in this set are labelled as follows: $\text{I}_{3a}$, $a=1,....,8$. The integral $\text{I}_{3a}$ is obtained from the integral $\text{I}_{2a}$, in set II, by performing the following replacements in the latter:
\begin{equation*}
\vec{k}_1\longleftrightarrow\vec{k}_3, \vec{k}_2\longleftrightarrow\vec{k}_4.
\end{equation*}

\subsection*{Set IV}

Furnished with the definition in (\ref{f2}), we introduce the following integrals
\begin{equation}
\begin{array}{l}
{
\text{I}_{41}=-\int_0^\infty\!\!d\omega\,\cfrac{\omega}{s^2(s+\omega^2)}\,f_{2}(\omega;k_1,k_2)\,f_{2}(\omega;k_3,k_4),}\\[2pt]
{
\text{I}_{42}=-\lim_{\epsilon\rightarrow 0}\int_0^\infty\!\!d\omega\,\cfrac{\omega}{s^2(s+\omega^2)}\;\cfrac{2(s^2-3\omega^4)}{((\sqrt{s}-\omega-i\epsilon)(\sqrt{s}+\omega-i\epsilon))^2}\,f_{2}(\omega;k_1,k_2)\,f_{2}(\omega;k_3,k_4),\quad
}\\[2pt]
{
\text{I}_{43}=\lim_{\epsilon\rightarrow 0}\int_0^\infty\!\!d\omega\,\cfrac{\omega}{s^2(s+\omega^2)}\;
\cfrac{4 s\omega^2}{((\sqrt{s}-\omega-i\epsilon)(\sqrt{s}+\omega-i\epsilon))^2}\,f_{2}(\omega;k_1,k_2)\,f_{2}(\omega;k_3,k_4),
}\\[2pt]
{
\text{I}_{44}=-\int_0^\infty\!\!d\omega\,\cfrac{\omega}{s^2(s+\omega^2)}\;\cfrac{s}{\omega^2}\,f_{2}(\omega;k_1,k_2)\,f_{2}(\omega;k_3,k_4),
}\\[2pt]
{
\text{I}_{45}=\lim_{\epsilon\rightarrow 0}\int_0^\infty\!\!d\omega\,\cfrac{\omega}{s^2(s+\omega^2)}\;
\cfrac{4 s^3}{\omega^2((\sqrt{s}-\omega-i\epsilon)(\sqrt{s}+\omega-i\epsilon))^2}\,f_{2}(\omega;k_1,k_2)\,f_{2}(\omega;k_3,k_4),
}\\[2pt]
{
\text{I}_{46}=\lim_{\epsilon\rightarrow 0}\int_0^\infty\!\!d\omega\,\cfrac{\omega}{s^2(s+\omega^2)}\;
\cfrac{-3 s^4+2s^3\omega^2+5s^2\omega^4}{((\sqrt{s}-\omega-i\epsilon)(\sqrt{s}+\omega-i\epsilon))^2}\,f_{2}(\omega;k_1,k_2)\,f_{2}(\omega;k_3,k_4),
}\\[2pt]
{
\text{I}_{47}=\lim_{\epsilon\rightarrow 0}\int_0^\infty\!\!d\omega\,
\cfrac{\omega(-s+3\omega^2)}{\omega^2((\sqrt{s}-\omega-i\epsilon)(\sqrt{s}+\omega-i\epsilon))^2}\,f_{2}(\omega;k_1,k_2)\,f_{2}(\omega;k_3,k_4),
}\\[2pt]
{
\text{I}_{48}=\lim_{\epsilon\rightarrow 0}\int_0^\infty\!\!d\omega\,
\cfrac{\omega(s+\omega^2)}{\omega^4((\sqrt{s}-\omega-i\epsilon)(\sqrt{s}+\omega-i\epsilon))^2}\,f_{2}(\omega;k_1,k_2)\,f_{2}(\omega;k_3,k_4),
}\\[2pt]
\label{int4}
\end{array}
\end{equation}
whose values read
\begin{equation}
\begin{array}{l}
{\text{I}_{41}=- \cfrac{A_{41}[k_1,k_2,k_3,k_4]}{(k_{12} + k_{34})^3 (k_{12} + \sqrt{s} )^2 (k_{34} + \sqrt{s} )^2 s^2}}
\label{i41}
\end{array}
\end{equation}
with
\begin{equation}
\begin{array}{l}
{A_{41}=k_1^4 (k_{34} + \sqrt{s})^2 + (k_2 + k_{34}) (k_2 + \sqrt{s}) \bigl[k_2 k_{34}(k_3^2 + k_2 k_{34} + 3 k_3 k_4 + k_4^2) +(2 k_2+  }\\[4pt]
{ +k_{34})(k_3^2 + k_2 k_{34} + 3 k_3 k_4 + k_4^2)\sqrt{s} + (k_2 + k_{34})^2 s\bigr]+ k_1^3 \Bigl[k_{34} \bigl(5 k_2 k_{34}+ (2 k_3 + k_4) (k_3 +}\\[4pt]
{+ 2 k_4)\bigr) + 2 (3 k_3^2 + 5 k_2 k_{34} + 7 k_3 k_4 + 3 k_4^2) \sqrt{s} + 5 (k_2 + k_{34}) s + \sqrt{s^3}\Bigr] + k_1^2 \Bigl[k_{34} \bigl(8 k_2^2 k_{34}  + }\\[4pt]
{ + k_{34} (k_3^2+ 3 k_3 k_4 + k_4^2)+ k_2 (7 k_3^2 + 19 k_3 k_4 + 7 k_4^2)\bigr) + 2 \bigl(k_3 (k_2 + k_3) (8 k_2 + 3 k_3) + (8 k_2^2 + }\\[4pt]
{+27 k_2 k_3 + 11 k_3^2) k_4 + 11 (k_2 + k_3) k_4^2 + 3 k_4^3\bigr)\sqrt{s} + 2 (4 k_2^2 + 4 k_3^2 + 9 k_2 k_{34} + 9 k_3 k_4 + 4 k_4^2) s + }\\[4pt]
{ +3 (k_2 + k_{34}) \sqrt{s^3}\Bigr] + k_1 \Bigl[k_2 k_{34} (5 k_2 + 2k_{34}) (k_3^2+ k_2 k_{34} + 3 k_3 k_4 + k_4^2) + 2 (k_2 + k_{34}) (5 k_2 +  }\\[4pt]
{+k_{34})(k_3^2 + k_2 k_{34} + 3 k_3 k_4 + k_4^2) \sqrt{s}+\bigl(5 k_2^3 + 18 k_2^2k_{34}+ 6 k_2 (3 k_3^2 + 7 k_3 k_4 + 3 k_4^2) + }\\[4pt]
{+k_{34} (5 k_3^2 + 13 k_3 k_4 + 5 k_4^2)\bigr) s + 3 (k_2 +k_{34})^2 \sqrt{s^3}\Bigr]}
\label{a41}
\end{array}
\end{equation}
\begin{equation*}
\begin{array}{l}
{\text{I}_{42}=- \cfrac{A_{42}[k_1,k_2,k_3,k_4]}{(k_{12} -  k_{34})^3 (k_{12} + k_{34})^3 s^2}}
\end{array}
\end{equation*}
with
\begin{equation*}
\begin{array}{l}
{A_{42}=- \cfrac{2 k_1 k_{12}^2 k_2 (k_{12}^2 -  k_{34}) (k_{12}^2 -  k_3^2 - 4 k_3 k_4 -  k_4^2)}{(k_{12}+ i \sqrt{s})^3} - \cfrac{k_{12} }{(k_{12} + i \sqrt{s})^2}\Bigl[k_1^2 \bigl(k_{12}^4 - 2 k_{12}^2 (k_3^2 +}\\[4pt]
{+ 3 k_3 k_4 + k_4^2)+ k_{34}^2 (k_3^2 + 4 k_3 k_4 + k_4^2)\bigr) + k_2^2 \bigl(k_{12}^4 - 2 k_{12}^2 (k_3^2  + 3 k_3 k_4 + k_4^2) + k_{34}^2 (k_3^2 + 4 k_3 k_4 +}\\[4pt]
{+ k_4^2)\bigr) + 4 k_4^2)\bigr)\Bigr]- \cfrac{4}{k_{12} + i \sqrt{s}} \Bigl[k_{12}^6 - 6 k_{12}^4 k_3 k_4 - 2 k_{12}^2 (k_1^2 + k_1 k_2 + k_2^2) (k_3^2 + k_4^2) + (k_1^2 + k_2^2) \times }\\[4pt]
{\times k_{34}^2 (k_3^2  + 4 k_3 k_4 + k_4^2)\Bigr]+\cfrac{2 k_3 k_{34}^2 (  k_{34}^2- k_{12}^2)  (- k_1^2 - 4 k_1 k_2 -  k_2^2 +k_{34}^2) k_4}{(k_{34} + i \sqrt{s})^3} + \cfrac{k_{34} }{(k_{34} + i \sqrt{s})^2}\times}\\[4pt]
{\times \Bigl[k_1^4 (k_3^2 + 3 k_3 k_4 + k_4^2)  + 6 k_1^3 k_2 (k_3^2 + 3 k_3 k_4+ k_4^2) + 2 k_1^2 \bigl(5 k_2^2 (k_3^2 + 3 k_3 k_4 + k_4^2) -  k_{34}^2 (k_3^2 +}\\[4pt]
{+ 4 k_3 k_4 + k_4^2)\bigr) + 6 k_1 k_2 \bigl(k_2^2 (k_3^2 + 3 k_3 k_4 + k_4^2) -  k_{34}^2 (k_3^2 + 5 k_3 k_4  +k_4^2)\bigr)+ (k_2^2 -  k_{34}^2)  \bigl(k_2^2 (k_3^2 + }\\[4pt]
{  +3 k_3 k_4 + k_4^2) -  k_{34}^2 (k_3^2 + 5 k_3 k_4 + k_4^2)\bigr)\Bigr]+ \cfrac{4 }{k_{34} + i \sqrt{s}}\Bigl[k_1^4 (k_3^2 + k_4^2)+ 6 k_1^3 k_2 (k_3^2 + k_4^2) + 6 k_1 k_2 \times }\\[4pt]
{\times \bigl(- k_{34}^4 + k_2^2 (k_3^2 + k_4^2)\bigr)  +(k_2^2 -  k_{34}^2)  \bigl(- k_{34}^4 + k_2^2 (k_3^2 + k_4^2)\bigr)+ 2 k_1^2 \bigl(5 k_2^2 (k_3^2 + k_4^2) -  k_{34}^2 (k_3^2 + }\\[4pt]
{+k_3 k_4 + k_4^2)\bigr)\Bigr] -  \cfrac{k_1 k_{12} k_2 (k_{12}^2-  k_{34}^2) (k_{12}^2 -  k_3^2 - 4 k_3 k_4 -  k_4^2)}{(k_{12} + \sqrt{s})^2} - \cfrac{1}{k_{12} + \sqrt{s}} \Big[k_1^6 + 6 k_1^5 k_2 +  }\\[4pt]
{+(6 k_1 k_2^3+k_2^2 (k_2^2 -  k_{34}^2))(k_2^2 -  k_3^2 - 4 k_3 k_4 -  k_4^2) + k_1^4 \bigl(15 k_2^2 - 2 (k_3^2 + 3 k_3 k_4 + k_4^2)\bigr)+ 2 k_1^3 k_2 \times }\\[4pt]
{\times \bigl(10 k_2^2 -3 (k_3^2  + 4 k_3 k_4+ k_4^2)\bigr) + k_1^2 \bigl(15 k_2^4 + k_{34}^2 (k_3^2 + 4 k_3 k_4 + k_4^2) - 4 k_2^2 (2 k_3^2 + 9 k_3 k_4 + 2 k_4^2)\bigr)\Big]+}\\[4pt]
{+\cfrac{k_3 k_{34}(k_{34}^2-k_{12}^2)(- k_1^2 - 4 k_1 k_2 -  k_2^2 +k_{34}^2) k_4}{(k_{34} +\sqrt{s})^2} +\cfrac{1}{k_{34}+\sqrt{s}}\Big[k_1^4 (k_3^2 + k_4^2)  + 6 k_1^3 k_2 (k_3^2 + k_4^2) +}\\[4pt]
{+ 6 k_1 k_2 \bigl(- k_{34}^4+  k_2^2 (k_3^2 + k_4^2)\bigr) +(k_2^2 -  k_{34}^2) \bigl(- k_{34}^4 + k_2^2 (k_3^2 + k_4^2)\bigr) + 2 k_1^2 \bigl(5 k_2^2 (k_3^2 + k_4^2) -}\\[4pt]
{-  k_{34}^2 (k_3^2 + k_3 k_4 + k_4^2)\bigr)\Big]}
\end{array}
\end{equation*}
\begin{equation*}
\begin{array}{l}
{\text{I}_{43}=- \cfrac{A_{43}[k_1,k_2,k_3,k_4]}{(k_{12} -  k_{34})^3 (k_{12} + k_{34})^3 s}}
\end{array}
\end{equation*}
with
\begin{equation*}
\begin{array}{l}
{A_{43}=\cfrac{2 k_1 k_2 (k_{12}^2 -  k_{34}^2) (k_{12}^2 -  k_3^2 - 4 k_3 k_4 -  k_4^2)}{(k_{12}+ i \sqrt{s})^3} + \cfrac{1}{k_{12} (k_{12} + i \sqrt{s})^2}\Big[k_1^6 + 9 k_1^5 k_2+ k_2^2 (k_2^2 -}\\[4pt]
{ -  k_{34}^2) (k_2^2-  k_3^2  - 4 k_3 k_4 -  k_4^2) + 3 k_1 k_2 (3 k_2^2 -  k_{34}^2) (k_2^2 -  k_3^2 - 4 k_3 k_4 -  k_4^2) + k_1^4 \bigl(27 k_2^2 - 2 (k_3^2 +}\\[4pt]
{+ 3 k_3 k_4 + k_4^2)\bigr) + 2 k_1^3 k_2 \bigl(19 k_2^2 - 3 (2 k_3^2 + 7 k_3 k_4 + 2 k_4^2)\bigr) + k_1^2 \bigl(27 k_2^4 + k_{34}^2 (k_3^2 + 4 k_3 k_4 + k_4^2) - }\\[4pt]
{-4 k_2^2 (5 k_3^2 +18 k_3 k_4 + 5 k_4^2)\bigr)\Big] +\cfrac{2 }{k_{12} + i\sqrt{s}} \bigl[k_1^4 + 6 k_1^3 k_2 + 6 k_1 k_2 (k_2^2 -  k_3^2 - 4 k_3 k_4 -  k_4^2) +}\\[4pt]
{+ (k_2^2 -  k_{34}^2)  (k_2^2  -  k_3^2- 4 k_3 k_4 -  k_4^2) + 2 k_1^2 (5 k_2^2 -  k_3^2 - 3 k_3 k_4 -  k_4^2)\bigr]+}\\[4pt]
{+ \cfrac{2 k_3 (k_{34}^2-k_{12}^2)  (k_1^2 + 4 k_1 k_2 + k_2^2 -  k_{34}^2) k_4}{(k_{34} + i\sqrt{s})^3}  - \cfrac{1}{k_{34} (k_{34} + i \sqrt{s})^2}\Big[k_1^4 (k_3^2+ 3 k_3 k_4 + k_4^2) +6 k_1^3 k_2\times }\\[4pt]
{\times (k_3^2 + 3 k_3 k_4 + k_4^2) + 2 k_1^2 \bigl(5 k_2^2 (k_3^2 + 3 k_3 k_4 + k_4^2) - k_{34}^2 (k_3^2 + 4 k_3 k_4 + k_4^2)\bigr) + 6 k_1 k_2 \bigl(k_2^2 (k_3^2 +}\\[4pt]
{+ 3 k_3 k_4 + k_4^2) -  k_{34}^2 (k_3^2 + 5 k_3 k_4 + k_4^2)\bigr) + (k_2^2 -k_{34}^2)  \bigl(k_2^2 (k_3^2 + 3 k_3 k_4 + k_4^2)-k_{34}^2 (k_3^2 + 5 k_3 k_4 + }\\[4pt]
{ +k_4^2)\bigr)\Big] - \cfrac{1}{k_{34} + i \sqrt{s}} \Big[2 (k_{12}^2 -  k_3^2)  (k_1^2 + 4 k_1 k_2 + k_2^2 -  k_3^2) - 12 k_3 (k_1^2 + 4 k_1 k_2+ k_2^2  -  k_3^2) k_4- }\\[4pt]
{ -4 (k_1^2 + 3 k_1 k_2 + k_2^2 - 5 k_3^2) k_4^2 + 12 k_3 k_4^3 + 2 k_4^4\Big]+\cfrac{k_1 k_2 (k_12 k_{12}^2-  k_{34}^2)(k_{12}^2 -  k_3^2 - 4 k_3 k_4 -  k_4^2)}{k_{12} (k_{12}+ \sqrt{s})^2} +  }\\[4pt]
{ + \cfrac{1}{k_{12} +\sqrt{s}}\Big[k_1^4 + 6 k_1^3 k_2 + 6 k_1 k_2 (k_2^2 -  k_3^2 - 4 k_3 k_4 -  k_4^2) + (k_2^2 - k_{34}^2)  (k_2^2 -  k_3^2 - 4 k_3 k_4 -  k_4^2)+ }\\[4pt]
{+2 k_1^2 (5 k_2^2 -  k_3^2 - 3 k_3 k_4 -  k_4^2)\Big] +\cfrac{k_3 (k_{34}^2-k_{12}^2)  (k_1^2 + 4 k_1 k_2 + k_2^2 -  k_{34}^2) k_4}{k_{34} (k_{34} + \sqrt{s})^2} -  }\\[4pt]
{ - \cfrac{1}{k_{34} + \sqrt{s}}\Big[k_1^4 + 6 k_1^3 k_2 + (k_2^2- k_{34}^2) (k_2^2 -  k_3^2 - 4 k_3 k_4 -  k_4^2) - 2 k_1^2 (-5 k_2^2 + k_3^2 + 3 k_3 k_4 + k_4^2) -}\\[4pt]
{- 6 k_1 k_2 (- k_2^2 + k_3^2 + 4 k_3 k_4 + k_4^2)\Big]}
\end{array}
\end{equation*}
\begin{equation*}
\begin{array}{l}
{\text{I}_{44}=- \cfrac{A_{44}[k_1,k_2,k_3,k_4]}{k_{12} k_{34} (k_{12} +k_{34})^3 s \bigl(k_{12}k_{34} + (k_{12} + k_{34}) \sqrt{s} + s\bigr)^2}}
\end{array}
\end{equation*}
with
\begin{equation*}
\begin{array}{l}
{A_{44}=k_1^4 (k_3^2 + 3 k_3 k_4 + k_4^2 + k_{34}\sqrt{s}) + k_2 (k_2 + k_{34}) (k_2 + \sqrt{s}) \bigl[k_2 k_3^2 + k_3^3+3 k_2 k_3 k_4  + }\\[4pt]
{ + 5 k_3^2 k_4+ k_2 k_4^2 + 5 k_3 k_4^2 + k_4^3 + (k_3^2 + k_2 k_{34} + 3 k_3 k_4 + k_4^2) \sqrt{s}\bigr]+k_1 \bigl(3 k_2 (2 k_2 +k_{34}) + }\\[4pt]
{ + (4 k_2 +k_{34}) \sqrt{s}\bigr) \bigl[k_2 k_3^2 + k_3^3 + 3 k_2 k_3 k_4 + 5 k_3^2 k_4 + k_2 k_4^2 + 5 k_3 k_4^2 + k_4^3 + (k_3^2 + k_2 k_{34} + }\\[4pt]
{ + 3 k_3 k_4 + k_4^2) \sqrt{s}\bigr]- k_1^3 \bigl[-2 k_3^3 - 9 k_3^2 k_4 - 9 k_3 k_4^2 - 2 k_4^3 - 6 k_2 (k_3^2 + 3 k_3 k_4 + k_4^2 + k_{34}\sqrt{s}) - }\\[4pt]
{-(3 k_3^2 + 8 k_3 k_4 + 3 k_4^2) \sqrt{s} -  k_{34} s\bigr] -  k_1^2 \Bigl[-10 k_2^2 (k_3^2 + 3 k_3 k_4 + k_4^2) -  k_{34}^2 (k_3^2 + 4 k_3 k_4 + k_4^2) - }\\[4pt]
{- 3 k_2 k_{34} (3 k_3^2 + 11 k_3 k_4 + 3 k_4^2) - \bigl(10 k_2^2 k_{34} + k_{34} (3 k_3 + k_4) (k_3 + 3 k_4) + k_2 (13 k_3^2 + }\\[4pt]
{+ 36 k_3 k_4 +13 k_4^2)\bigr) \sqrt{s} - \bigl(4 k_2 k_{34} + (2 k_3 + k_4) (k_3 + 2 k_4)\bigr) s\Bigr]}
\end{array}
\end{equation*}
\begin{equation*}
\begin{array}{l}
{\text{I}_{45}=\cfrac{2 k_1 k_2 (k_{12}^2 -  k_3^2 - 4 k_3 k_4 -  k_4^2)}{k_{12}^2 (k_{12}^2 -  k_{34}^2)^2  (k_{12} + i \sqrt{s})^3} + \cfrac{1}{k_{12}^3 (k_{12}^2 -  k_{34}^2)^3  (k_{12} + i\sqrt{s})^2}\Big[k_1^6 + 9 k_1^5 k_2 + k_2^2 (k_2^2 -  k_{34}^2)\times}\\[4pt]
{\times  (k_2^2 -  k_3^2 - 4 k_3 k_4 -  k_4^2) + 3 k_1 k_2 (3 k_2^2 -  k_{34}^2) (k_2^2 -  k_3^2 - 4 k_3 k_4 -  k_4^2) + k_1^4 \bigl(27 k_2^2 - 2 (k_3^2 + 3 k_3 k_4 +}\\[4pt]
{+ k_4^2)\bigr) + 2 k_1^3 k_2 \bigl(19 k_2^2 - 3 (2 k_3^2 + 7 k_3 k_4 + 2 k_4^2)\bigr) + k_1^2 \bigl(27 k_2^4 + k_{34}^2 (k_3^2 + 4 k_3 k_4 + k_4^2) -4 k_2^2 (5 k_3^2+}\\[4pt]
{ + 18 k_3 k_4 + 5 k_4^2)\bigr)\Big] -  \cfrac{2 k_3 (k_1^2 + 4 k_1 k_2 + k_2^2 -  k_{34}^2) k_4}{(k_{12}^2 - k_{34}^2)^2 k_{34}^2 (k_{34} + i \sqrt{s})^3} -  \cfrac{1}{(k_{12}^2 -  k_{34}^2)^3k_{34}^3 (k_{34} + i \sqrt{s})^2} \Big[k_1^4 (k_3^2   +}\\[4pt]
{+ 3 k_3 k_4 + k_4^2)+ 6 k_1^3 k_2 (k_3^2 + 3 k_3 k_4 + k_4^2)+ 2 k_1^2 \bigl(5 k_2^2 (k_3^2 + 3 k_3 k_4 + k_4^2) -  k_{34}^2 (k_3^2 + 4 k_3 k_4 + k_4^2)\bigr)+}\\[4pt]
{ + 6 k_1 k_2 \bigl(k_2^2 (k_3^2 + 3 k_3 k_4 + k_4^2) - k_{34}^2 (k_3^2 + 5 k_3 k_4 + k_4^2)\bigr) +(k_2^2 -  k_{34}^2)  \bigl(k_2^2 (k_3^2 + 3 k_3 k_4+ k_4^2)-}\\[4pt]
{  -  k_{34}^2 (k_3^2 + 5 k_3 k_4 + k_4^2)\bigr)\Big]+ \cfrac{k_1 k_2 (- k_{12}^2 + k_3^2 + 4 k_3 k_4 + k_4^2)}{k_{312}^3 (k_{12}^2 -  k_{34}^2)^2 (k_{12} + \sqrt{s})^2} -  \cfrac{1}{k_{12}^4 (k_{12}^2 -  k_{34}^2)^3 (k_{12} + \sqrt{s})} \times}\\[4pt]
{\times \Big[k_1^6  + 10 k_1^5 k_2+ 2 k_1^3 k_2 (22 k_2^2 - 7 k_3^2 - 24 k_3 k_4 - 7 k_4^2) + k_2^2 (k_2^2 - k_{34}^2)  (k_2^2 -  k_3^2 - 4 k_3 k_4 -  k_4^2) +}\\[4pt]
{+ 2 k_1 k_2 (5 k_2^2 - 2 k_{34}^2) (k_2^2 -  k_3^2 - 4 k_3 k_4 -  k_4^2) + k_1^4 \bigl(31 k_2^2 - 2 (k_3^2 + 3 k_3 k_4 + k_4^2)\bigr) +}\\[4pt]
{+ k_1^2 \bigl(31 k_2^4 + k_{34}^2 (k_3^2 + 4 k_3 k_4 + k_4^2) - 12 k_2^2 (2 k_3^2 + 7 k_3 k_4 + 2 k_4^2)\bigr)\Big]+ \cfrac{k_3 (k_1^2 + 4 k_1 k_2 + k_2^2 -  k_{34}^2) k_4}{(k_{12}^2 -  k_{34}^2)^2 k_{34}^3 (k_{34} + \sqrt{s})^2} +}\\[4pt]
{+ \cfrac{1}{(k_{12}^2 -  k_{34}^2)^3 k_{34}^4  (k_{34} +\sqrt{s})}\Big[k_1^4 (k_3^2 + 4 k_3 k_4 + k_4^2) + 6 k_1^3 k_2 (k_3^2 + 4 k_3 k_4 + k_4^2) + 2 k_1^2 \bigl(5 k_2^2 (k_3^2  +}\\[4pt]
{+ 4 k_3 k_4+ k_4^2) -  k_{34}^2 (k_3^2 + 5 k_3 k_4 + k_4^2)\bigr) + 6 k_1 k_2 \bigl(k_2^2 (k_3^2 + 4 k_3 k_4 + k_4^2) - k_{34}^2 (k_3^2 + 6 k_3 k_4 + k_4^2)\bigr) + }\\[4pt]
{+(k_2^2 -  k_{34}^2)\bigl(k_2^2 (k_3^2 + 4 k_3 k_4 + k_4^2) -  k_{34}^2 (k_3^2 + 6 k_3 k_4 + k_4^2)\bigr)\Big] - }\\[4pt]
{- \cfrac{(k_1^2 + 4 k_1 k_2 + k_2^2) (k_3^2 + 4 k_3 k_4 + k_4^2)}{k_{12}^4 k_{34}^4\sqrt{s}}}
\end{array}
\end{equation*}
\begin{equation*}
\begin{array}{l}
{\text{I}_{46}=\cfrac{A_{46}[k_1,k_2,k_3,k_4]}{(k_{12} + k_{34})^3 (k_{12} + i \sqrt{s})^3 (k_{34} + i\sqrt{s})^3}}
\end{array}
\end{equation*}
with
\begin{equation*}
\begin{array}{l}
{A_{46}=5 k_1^5 (k_{34} + i \sqrt{s})^3 + k_1^4 \Big[5 k_{34}^2 (2 k_3 + k_4) (k_3 + 2 k_4) + 30 k_2 (k_{34} + i \sqrt{s})^3 + }\\[4pt]
{+15i  (3 k_3^2 + 7 k_3 k_4 + 3 k_4^2) \sqrt{s} - 4 (17 k_3^2 + 36 k_3 k_4 + 17 k_4^2) s - 42i k_{34}\sqrt{s^3} + 9 s^2\Big] +}\\[4pt]
{+ (k_2 + k_{34}) (k_2 + i \sqrt{s}) \Big[5 k_2^2 k_{34}^2 (k_3^2 + k_2 k_{34} + 3 k_3 k_4 + k_4^2) + 5i k_2 k_{34} (3 k_2 + 2 k_{34})\times  }\\[4pt]
{\times(k_3^2 + k_2 k_{34} + 3 k_3 k_4 + k_4^2) \sqrt{s} - \bigl(15 k_2^3k_{34} + 5k_{34}^2 (k_3^2 + 3 k_3 k_4 + k_4^2) + 4 k_2k_{34} (7 k_3^2 +}\\[4pt]
{+ 18 k_3 k_4 + 7 k_4^2) + k_2^2 (38 k_3^2 + 84 k_3 k_4 + 38 k_4^2)\bigr) s - i \bigl(5 k_2^3 + 26 k_2 k_3^2 + 22 k_2^2 k_{34}  +}\\[4pt]
{+ 9 k_3^2 k_{34}+ k_3 (58 k_2 + 21 k_{34}) k_4 + (26 k_2 + 9 k_{34}) k_4^2\bigr) \sqrt{s^3} + 4 (k_2 + k_{34})^2 s^2\Big] + }\\[4pt]
{+3 k_1 (2 k_2 + k_{34} + i \sqrt{s}) \Big[5 k_2^2 k_{34}^2 (k_3^2 + k_2 k_{34} + 3 k_3 k_4 + k_4^2) +5i k_2k_{34}(3 k_2+ 2k_{34}) (k_3^2 + }\\[4pt]
{ + k_2 k_{34} + 3 k_3 k_4 + k_4^2) \sqrt{s} -\bigl(15 k_2^3 k_{34} + 5 k_{34}^2 (k_3^2 + 3 k_3 k_4 + k_4^2) + 4 k_2 k_{34} (7 k_3^2 + 18 k_3 k_4+}\\[4pt]
{ + 7 k_4^2) + k_2^2 (38 k_3^2 + 84 k_3 k_4 + 38 k_4^2)\bigr) s - i \bigl(5 k_2^3 + 26 k_2 k_3^2 + 22 k_2^2 k_{34} + 9 k_3^2 k_{34} + k_3 (58 k_2+}\\[4pt]
{  + 21 k_{34}) k_4+ (26 k_2 + 9 k_{34}) k_4^2\bigr) \sqrt{s^3} + 4 (k_2 + k_{34})^2 s^2\Big] + k_1^3 \Big[5 k_{34}^2 \bigl(9 k_2 k_3^2 + 13 k_2^2k_{34}+}\\[4pt]
{ + k_3^2 k_{34} + 3 k_3 (8 k_2 + k_{34}) k_4 + (9 k_2 + k_{34}) k_4^2\bigr) +15i k_{34} \bigl(13 k_2^2 k_{34} + k_{34} (3 k_3^2 + 8 k_3 k_4 + }\\[4pt]
{ + 3 k_4^2)+ 2 k_2 (7 k_3^2 + 17 k_3 k_4 + 7 k_4^2)\bigr) \sqrt{s}-  \bigl(318 k_2 k_3^2 + 195 k_2^2 k_{34}+ 106 k_3^2 k_{34} + k_3 (684 k_2 +}\\[4pt]
{+ 251 k_{34}) k_4 + 106 (3 k_2 + k_{34}) k_4^2\bigr) s - i (65 k_2^2 + 101 k_3^2 + 192 k_2 k_{34} + 216 k_3 k_4 + 101 k_4^2) \sqrt{s^3} +}\\[4pt]
{+ 39 (k_2 + k_{34}) s^2 + 4i \sqrt{s^5}\Big] + k_1^2 \Big[5 k_2 k_{34}^2 \bigl(13 k_2^2 k_{34} + 3 k_{34} (k_3^2 + 3 k_3 k_4 + k_4^2) + 2 k_2 (7 k_3^2+ }\\[4pt]
{ + 19 k_3 k_4 + 7 k_4^2)\bigr) +15i k_{34} \bigl(13 k_2^3 k_{34} + k_{34}^2 (k_3^2 + 3 k_3 k_4 + k_4^2) + 2 k_2 k_{34} (5 k_3^2 + 14 k_3 k_4 +}\\[4pt]
{+ 5 k_4^2) + k_2^2 (22 k_3^2 + 54 k_3 k_4 + 22 k_4^2)\bigr) \sqrt{s} -  \bigl(195 k_2^3 k_{34} + 20 k_2^2 (25 k_3^2 + 54 k_3 k_4 + 25 k_4^2) +}\\[4pt]
{+ 2 k_{34}^2 (34 k_3^2 + 91 k_3 k_4 + 34 k_4^2) + 3 k_2 k_{34} (119 k_3^2 + 293 k_3 k_4 + 119 k_4^2)\bigr) s - i \bigl(65 k_2^3 +}\\[4pt]
{ + 300 k_2^2 k_{34} + 12 k_2 (28 k_3^2 + 61 k_3 k_4 + 28 k_4^2)+ k_{34} (101 k_3^2 + 235 k_3 k_4 + 101 k_4^2)\bigr) \sqrt{s^3} +}\\[4pt]
{+ 6 (10 k_2^2 + 10 k_3^2 + 21 k_2 k_{34} + 21 k_3 k_4 + 10 k_4^2) s^2 + 12i (k_2 + k_{34}) \sqrt{s^5}\Big]}
\end{array}
\end{equation*}
\begin{equation*}
\begin{array}{l}
{\text{I}_{47}=\cfrac{A_{47}[k_1,k_2,k_3,k_4]}{k_{12}k_{34}(k_{12} + k_{34})^3 (k_{12} + i \sqrt{s})^3 (k_{34} + i \sqrt{s})^3}}
\end{array}
\end{equation*}
with
\begin{equation*}
\begin{array}{l}
{A_{47}=k_1^5 \bigl(3 k_{34} (k_3^2 + 3 k_3 k_4 + k_4^2) + i (5 k_3^2 + 11 k_3 k_4 + 5 k_4^2) \sqrt{s} - 2 k_{34} s\bigr)+ k_1 \bigl(k_2 (7 k_2  +}\\[4pt]
{+ 4 k_{34}) + i (4 k_2 + k_{34}) \sqrt{s} \bigr) \Bigl(3 k_2 k_{34} \bigl(k_3^2 (k_2 + k_{34}) + k_3 (3 k_2 + 4 k_{34}) k_4 +(k_2 + k_{34}) k_4^2\bigr)+ }\\[4pt]
{ + i \bigl(k_3^2 (k_2 + k_{34}) (5 k_2 + 2k_{34}) + k_3 (11 k_2^2 + 22 k_2 k_{34} + 8 k_{34}^2) k_4 + (k_2 +k_{34}) (5 k_2 + }\\[4pt]
{+ 2 k_{34}) k_4^2\bigr) \sqrt{s} -  (2 k_2 + 3k_{34}) (k_3^2 + k_2k_{34} + 3 k_3 k_4 + k_4^2) s - i (k_3^2 + k_2k_{34}+ 3 k_3 k_4 + }\\[4pt]
{+k_4^2) \sqrt{s^3} \Bigr) + k_1^3 \Bigl(3 k_{34} \bigl(16 k_2^2 (k_3^2 + 3 k_3 k_4 + k_4^2) + k_{34}^2 (k_3^2 + 4 k_3 k_4 + k_4^2) + k_2 k_{34} (11 k_3^2+}\\[4pt]
{ + 40 k_3 k_4 + 11 k_4^2)\bigr) + i \bigl(16 k_2^2 (5 k_3^2 + 11 k_3 k_4 + 5 k_4^2)+3 k_{34}^2 (5 k_3^2 + 17 k_3 k_4 + 5 k_4^2) +  }\\[4pt]
{ + 9 k_2k_{34} (9 k_3^2 +26 k_3 k_4 + 9 k_4^2)\bigr) \sqrt{s} - \bigl(63 k_2 k_3^2 + 32 k_2^2 k_{34} + 20 k_3^2k_{34}+ 18 k_3 (8 k_2 +  }\\[4pt]
{+3k_{34}) k_4 + (63 k_2 + 20k_{34}) k_4^2\bigr) s - 3i (3 k_3^2 + 5 k_2 k_{34} + 7 k_3 k_4 + 3 k_4^2) \sqrt{s^3} + k_{34} s^2\Bigr) +}\\[4pt]
{+ k_2 (k_2 + k_{34}) \Bigl(3 k_2^2 k_{34} \bigl(k_3^2 (k_2 + k_{34}) + k_3 (3 k_2 + 4k_{34}) k_4 + (k_2 + k_{34}) k_4^2\bigr)+ i k_2 (k_2 +}\\[4pt]
{ + k_{34}) \bigl(5 k_3^2 (k_2 +k_{34}) + k_3 (11 k_2 + 20k_{34}) k_4 + 5 (k_2 +k_{34}) k_4^2\bigr) \sqrt{s}  -\bigl(2 (5 k_2^2 k_3^2+ k_2^3 k_{34}+  }\\[4pt]
{  + 5 k_2 k_3^2 k_{34} + k_3^2 k_{34}^2) + k_3 (k_2 + k_{34}) (23 k_2 + 8 k_{34}) k_4 + 2 (5 k_2^2 + 5 k_2 k_{34} + k_{34}^2) k_4^2\bigr) s -}\\[4pt]
{- 3i (k_2 + k_{34}) (k_3^2 + k_2 k_{34} + 3 k_3 k_4 + k_4^2) \sqrt{s^3}  + (k_3^2 + k_2k_{34} + 3 k_3 k_4 + k_4^2) s^2\Bigr) +}\\[4pt]
{+ k_1^2 \Bigl[6 k_2 k_{34} \bigl(8 k_2^2 (k_3^2 + 3 k_3 k_4 + k_4^2) + 2 k_{34}^2 (k_3^2 + 4 k_3 k_4 + k_4^2)+ 3 k_2k_{34}(3 k_3^2 + 11 k_3 k_4+}\\[4pt]
{  + 3 k_4^2)\bigr) + i \bigl(k_3^2 (k_2 + k_{34}) (80 k_2^2 + 52 k_2 k_{34} + 5 k_{34}^2)+ k_3 (176 k_2^3 + 384 k_2^2 k_{34} + 201 k_2 k_{34}^2 +}\\[4pt]
{ + 20 k_{34}^3) k_4 + (k_2 + k_{34}) (80 k_2^2 + 52 k_2 k_{34} + 5 k_{34}^2) k_4^2\bigr) \sqrt{s}-\bigl(32 k_2^3k_{34}+ 3k_{34}^2 (4 k_3^2 + }\\[4pt]
{+ 13 k_3 k_4 + 4 k_4^2) + 6 k_2^2 (17 k_3^2 + 39 k_3 k_4 + 17 k_4^2) + 2 k_2k_{34}(37 k_3^2 + 105 k_3 k_4 + 37 k_4^2)\bigr) s-}\\[4pt]
{ - 3i \bigl(k_3 (k_2 + k_3) (8 k_2 + 3 k_3) +(8 k_2^2 + 27 k_2 k_3 + 11 k_3^2) k_4 + 11 (k_2 + k_3) k_4^2 + 3 k_4^3\bigr) \sqrt{s^3} + }\\[4pt]
{+\bigl(4 k_2 k_{34} + (2 k_3 + k_4) (k_3 + 2 k_4)\bigr) s^2\Bigr]+ k_1^4 \Bigl[7 k_2 \bigl(3 k_{34} (k_3^2 + 3 k_3 k_4 + k_4^2) + i (5 k_3^2  +}\\[4pt]
{+ 11 k_3 k_4 + 5 k_4^2) \sqrt{s}- 2k_{34} s\bigr) - 3 (k_{34} + i \sqrt{s} ) \bigl(-2 k_3^3 - 9 k_3^2 k_4 - 9 k_3 k_4^2 - 2 k_4^3 -}\\[4pt]
{- i (3 k_3^2 + 7 k_3 k_4 + 3 k_4^2) \sqrt{s}  + k_{34} s\bigr)\Bigr]}
\end{array}
\end{equation*}
\begin{equation*}
\begin{array}{l}
{\text{I}_{48}=\cfrac{A_{48}[k_1,k_2,k_3,k_4]}{k_{12}^3k_{34}^3(k_{12} + k_{34})^3 (k_{12} + i \sqrt{s})^3 (k_{34} + i \sqrt{s})^3}
}
\end{array}
\end{equation*}
with
\begin{equation*}
\begin{array}{l}
{A_{48}=k_1^7 \bigl(k_3^3 + 6 k_3^2 k_4 + 6 k_3 k_4^2 + k_4^3 + i (k_3^2 + 3 k_3 k_4 + k_4^2) \sqrt{s}\bigr)+ 3 k_1^6 (3 k_2 + k_{34} +}\\[4pt]
{ + i \sqrt{s}) \bigl(k_3^3 + 6 k_3^2 k_4 + 6 k_3 k_4^2 + k_4^3 + i (k_3^2 + 3 k_3 k_4 + k_4^2) \sqrt{s}\bigr)+ k_1^5 \Bigl[k_{34} \bigl(31 k_2^2 (k_3^2  +}\\[4pt]
{+ 5 k_3 k_4 + k_4^2) + 24 k_2 k_{34} (k_3^2 + 5 k_3 k_4 + k_4^2) + 2 k_{34}^2 (2 k_3^2 + 9 k_3 k_4 + 2 k_4^2)\bigr) + i \bigl(31 k_2^2 (k_3^2 + }\\[4pt]
{+ 3 k_3 k_4 + k_4^2)  + 48 k_2 k_{34}(k_3^2 + 4 k_3 k_4 + k_4^2)+k_{34}^2 (13 k_3^2 + 55 k_3 k_4 + 13 k_4^2)\bigr) \sqrt{s}-}\\[4pt]
{- 6 \bigl(4 k_2 (k_3^2 + 3 k_3 k_4 + k_4^2) +k_{34} (2 k_3^2 + 7 k_3 k_4 + 2 k_4^2)\bigr) s-3i (k_3^2 + 3 k_3 k_4 + k_4^2) \sqrt{s^3}\Bigr] +}\\[4pt]
{+k_1^4 \Bigl[k_{34} \bigl(3 k_{34}^3 (k_3^2 + 4 k_3 k_4 + k_4^2) + 55 k_2^3 (k_3^2 + 5 k_3 k_4 + k_4^2) + 69 k_2^2 k_{34} (k_3^2 + 5 k_3 k_4 +}\\[4pt]
{+ k_4^2) + 6 k_2 k_{34}^2 (5 k_3^2 + 22 k_3 k_4 + 5 k_4^2)\bigr) +i \bigl(55 k_2^3 (k_3^2 + 3 k_3 k_4 + k_4^2) + 138 k_2^2 k_{34} (k_3^2 +}\\[4pt]
{+ 4 k_3 k_4 + k_4^2) + k_{34}^3 (13 k_3^2 + 54 k_3 k_4 + 13 k_4^2) + 3 k_2k_{34}^2 (31 k_3^2 + 129 k_3 k_4 + 31 k_4^2)\bigr)\sqrt{s}- }\\[4pt]
{-3 \bigl(23 k_2^2 (k_3^2 + 3 k_3 k_4 + k_4^2) + 14 k_2 k_{34} (2 k_3^2 + 7 k_3 k_4 + 2 k_4^2) + k_{34}^2 (6 k_3^2 + 23 k_3 k_4 + 6 k_4^2)\bigr) s-}\\[4pt]
{ - 3i \bigl(k_{34} (3 k_3 + k_4) (k_3 + 3 k_4) +7 k_2 (k_3^2 + 3 k_3 k_4 + k_4^2)\bigr) \sqrt{s^3} + (k_3^2 + 3 k_3 k_4 + k_4^2) s^2\Bigr] +}\\[4pt]
{+ k_2^2 (k_2 + k_{34}) (k_2 + i \sqrt{s}) \Bigl(k_{34} (k_3^2 + 4 k_3 k_4 + k_4^2) (k_{34} + i \sqrt{s})^3+k_2^3 \bigl[k_3^3 + 6 k_3^2 k_4 + }\\[4pt]
{ + 6 k_3 k_4^2 + k_4^3 + i (k_3^2 + 3 k_3 k_4 + k_4^2)\sqrt{s}\bigr] + 2 k_2^2 \bigl[k_{34}^2 (k_3^2 + 5 k_3 k_4 + k_4^2) + 2i k_{34} (k_3^2+}\\[4pt]
{ + 4 k_3 k_4 + k_4^2) \sqrt{s} -  (k_3^2 + 3 k_3 k_4 + k_4^2) s\bigr] + k_2 \bigl[2 k_{34}^3 (k_3^2 + 4 k_3 k_4 + k_4^2)+ 6i k_{34}^2 (k_3^2  +}\\[4pt]
{+ 4 k_3 k_4 + k_4^2) \sqrt{s}-  k_{34} (5 k_3^2 + 17 k_3 k_4 + 5 k_4^2) s - i (k_3^2 + 3 k_3 k_4 + k_4^2) \sqrt{s^3}\bigr]\Bigr) +}\\[4pt]
{+ 3 k_1 k_2 \bigl(k_2 (3 k_2 + 2 k_{34}) + i (2 k_2 + k_{34}) \sqrt{s}\bigr) \Bigl(k_{34} (k_3^2 + 4 k_3 k_4 + k_4^2) (k_{34} + i \sqrt{s})^3 +}\\[4pt]
{+ k_2^3 \bigl(k_3^3 + 6 k_3^2 k_4 + 6 k_3 k_4^2 + k_4^3 + i (k_3^2 + 3 k_3 k_4 + k_4^2) \sqrt{s}\bigr)+2 k_2^2 \bigl(k_{34}^2 (k_3^2 + 5 k_3 k_4 + k_4^2) + }\\[4pt]
{ + 2i k_{34} (k_3^2 + 4 k_3 k_4 + k_4^2) \sqrt{s}- (k_3^2 + 3 k_3 k_4 + k_4^2) s\bigr) + k_2 \bigl(2 k_{34}^3 (k_3^2 + 4 k_3 k_4 + k_4^2) +}\\[4pt]
{+ 6i k_{34}^2 (k_3^2 + 4 k_3 k_4 + k_4^2) \sqrt{s} -  k_{34} (5 k_3^2 + 17 k_3 k_4 + 5 k_4^2) s - i (k_3^2 + 3 k_3 k_4 + k_4^2) \sqrt{s^3}\bigr)\Bigr) + }\\[4pt]
{+k_1^3 \Bigl[55 k_2^4 \bigl(k_3^3 + 6 k_3^2 k_4 + 6 k_3 k_4^2 + k_4^3 + i (k_3^2 + 3 k_3 k_4 + k_4^2) \sqrt{s}\bigr) + 96 k_2^3 \bigl(k_{34}^2 (k_3^2 + 5 k_3 k_4+}\\[4pt]
{ + k_4^2) + 2i k_{34} (k_3^2 + 4 k_3 k_4 + k_4^2) \sqrt{s} -  (k_3^2 + 3 k_3 k_4 + k_4^2) s\bigr) + 2 k_2^2 \bigl(k_{34}^3 (35 k_3^2 + 153 k_3 k_4 +}\\[4pt]
{+ 35 k_4^2) + i k_{34}^2 (107 k_3^2 + 443 k_3 k_4 + 107 k_4^2) \sqrt{s} - 48 k_{34} (2 k_3^2 + 7 k_3 k_4 + 2 k_4^2) s - 24i (k_3^2 +}\\[4pt]
{+ 3 k_3 k_4 + k_4^2) \sqrt{s^3}\bigr) + 3 k_2 \bigl(7 k_{34}^4 (k_3^2 + 4 k_3 k_4 + k_4^2) + i k_{34}^3 (27 k_3^2 + 110 k_3 k_4 + 27 k_4^2) \sqrt{s} - }\\[4pt]
{- k_{34}^2 (35 k_3^2 + 134 k_3 k_4 + 35 k_4^2) s - i k_{34} (17 k_3^2 + 58 k_3 k_4 + 17 k_4^2) \sqrt{s^3}+ 2 (k_3^2 + 3 k_3 k_4 +}\\[4pt]
{ +k_4^2) s^2\bigr)  + k_{34}\bigl(k_{34}^4 (k_3^2 + 4 k_3 k_4 + k_4^2) + 6i k_{34}^3 (k_3^2 + 4 k_3 k_4 + k_4^2) \sqrt{s} - 12 k_{34}^2 (k_3^2 + 4 k_3 k_4 + }\\[4pt]
{+k_4^2) s - 3i k_{34} (3 k_3^2 + 11 k_3 k_4 + 3 k_4^2) \sqrt{s^3} + (2 k_3^2 + 7 k_3 k_4 + 2 k_4^2) s^2\bigr)\Bigr]+ }\nonumber
\end{array}
\end{equation*}
\raggedbottom
\begin{equation*}
\begin{array}{l}
{+ k_1^2 \Bigl[31 k_2^5 \bigl(k_3^3 + 6 k_3^2 k_4 + 6 k_3 k_4^2 + k_4^3 + i (k_3^2 + 3 k_3 k_4 + k_4^2) \sqrt{s}\bigr)+ i k_{34}^2 (k_3^2 +}\\[4pt]
{ + 4 k_3 k_4 + k_4^2) (k_{34} + i \sqrt{s})^3 \sqrt{s} + 69 k_2^4 \bigl(k_{34}^2 (k_3^2 + 5 k_3 k_4 + k_4^2) + 2i k_{34} (k_3^2 + 4 k_3 k_4 + }\\[4pt]
{+k_4^2) \sqrt{s} -  (k_3^2 + 3 k_3 k_4 + k_4^2) s\bigr) +2 k_2^3 \bigl(k_{34}^3 (35 k_3^2 + 153 k_3 k_4 + 35 k_4^2) + i k_{34}^2 (107 k_3^2+ }\\[4pt]
{ + 443 k_3 k_4 + 107 k_4^2) \sqrt{s} -48 k_{34} (2 k_3^2 + 7 k_3 k_4 + 2 k_4^2) s - 24i (k_3^2 + 3 k_3 k_4 + k_4^2) \sqrt{s^3}\bigr) +}\\[4pt]
{+ 2 k_2^2 \bigl(18 k_{34}^4 (k_3^2 + 4 k_3 k_4 + k_4^2) + 4i k_{34}^3 (17 k_3^2 + 69 k_3 k_4 + 17 k_4^2) \sqrt{s} - 3 k_{34}^2 (29 k_3^2 +}\\[4pt]
{+ 111 k_3 k_4 + 29 k_4^2) s - 6i k_{34} (7 k_3^2 + 24 k_3 k_4 + 7 k_4^2) \sqrt{s^3} + 5 (k_3^2 + 3 k_3 k_4 + k_4^2) s^2\bigr) +}\\[4pt]
{+ 3 k_2 k_{34} \bigl(2 k_{34}^4 (k_3^2 + 4 k_3 k_4 + k_4^2) + 10i k_{34}^3 (k_3^2 + 4 k_3 k_4 + k_4^2) \sqrt{s} -18 k_{34}^2 (k_3^2 + 4 k_3 k_4 + }\\[4pt]
{+ k_4^2) s - i k_{34} (13 k_3^2 + 49 k_3 k_4 + 13 k_4^2) \sqrt{s^3} + (3 k_3^2 + 11 k_3 k_4 + 3 k_4^2) s^2\bigr)\Bigr]}\\[4pt]
\end{array}
\end{equation*}

\section{The $s$-channel graviton exchange in General Relativity.}

The mathematical object, ${\cal W}^{(GR)}_s$,  corresponding to the Witten diagram in Figure 1 is, for General Relativity, given in (\ref{WGRsz1z2}). Taking into account the splitting in three terms of the General Relativity bulk-to-bulk propagator  in (\ref{BtBpropGR}), one concludes that
\begin{equation}
{\cal W}^{(GR)}_s=(2\pi)^3\delta\big(\sum_{a=1}^4\vec{k}_a\big)\kappa^2\big({\cal W}^{(1)}_s+{\cal W}^{(2)}_s+{\cal W}^{(3)}_s\big),
\label{WGRtotal}
\end{equation}
where
\begin{equation}
\begin{array}{l}
{
{\cal W}^{(c)}_s\!\!=\!\!(2\pi)^3\!\delta\big(\sum_{a=1}^4\vec{k}_a\big)\kappa^2\int_{0}^\infty\!\! dz_1\!\!\int_{0}^\infty\!\! dz_2\; V_L^{(GR)\,i_1 j_1}(z_1;k_1,k_2)\,G^{(c)}_{i_1 j_2,i_2 j_2}(z_1,z_ 2;\vec{k})\, V_R^{(GR)\,i_1 j_1}(z_2;k_3,k_4),}\\[8pt]
{
G^{(c)}_{i_1 j_2,i_2 j_2}(z_1,z_ 2;\vec{k})=(z_1 z_ 2)^{-1/2}\int_{0}^{\infty} d\omega\,\omega\, J_{3/2}[\omega z_1]\,{\tilde G}^{(c)}_{i_1 j_1,i_2 j_2}(\omega,\vec{k})\,J_{3/2}[\omega z_2],\quad c=1,2,3,}\\[8pt]
{
{\tilde G}^{(1)}_{i_1 j_1,i_2 j_2}(\omega,\vec{k})= G_1^{(GR)}(T_{i_1 i_2} T_{j_1 j_2}+T_{i_1 j_2} T_{j_1 i_2}-T_{i_1 j_1} T_{i_2 j_2}),
}\\[4pt]
{
{\tilde G}^{(2)}_{i_1 j_1,i_2 j_2}(\omega,\vec{k})=
G_2^{(GR)}(T_{i_1 i_2} L_{j_1 j_2}+L_{i_1 i_2} T_{j_1 j_2}+T_{i_1 j_2} L_{j_1 i_2}+L_{i_1 j_2} T_{j_1 i_2}  -T_{i_1 j_1} L_{i_2 j_2}-L_{i_1 j_1} T_{i_2 j_2}+}\\[2pt]
{\phantom{{\tilde G}^{(2)}_{i_1 j_1,i_2 j_2}(\omega,\vec{k})=G_2^{(GR)}(}
L_{i_1 i_2} L_{j_1 j_2}  +L_{i_1 j_2} L_{j_1 i_2}-L_{i_1 j_1} L_{i_2 j_2}),
}\\[8pt]
{
{\tilde G}^{(3)}_{i_1 j_1,i_2 j_2}(\omega,\vec{k})=
 G_3^{(GR)}(L_{i_1 i_2} L_{j_1 j_2}+L_{i_1 j_2} L_{j_1 i_2}-L_{i_1 j_1} L_{i_2 j_2}).
}
\label{WGR123def}
\end{array}
\end{equation}
The definitions of $T_{ij}$, $L_{ij}$, $ G_1^{(GR)}$,  $G_2^{(GR)}$ and $ G_3^{(GR)}$ can be found in (\ref{coefBtBpropGR}).

After doing the tensor contractions involved in the computation of ${\cal W}^{(1)}_s$, one gets
\begin{equation}
{\cal W}^{(1)}_s=\cfrac{L^2}{2}\, N_{11}\, \text{I}_{41},
\label{W1value}
\end{equation}
where $\text{I}_{41}$ is given in (\ref{int4}), (\ref{i41}) and (\ref{a41}), and
\begin{equation}
\begin{array}{l}
{N_{11}=\cfrac{1}{16}\big[ k_2^4 (k_3^4 - 2 k_3^2 (k_4^2 - 3 s) + (k_4^2 - s)^2) +
   k_1^4 (k_3^4 + k_4^4 + 6 k_4^2 s + s^2 - 2 k_3^2 (k_4^2 + s)) +}\\[4pt]
{\quad   s^2 (k_3^4 + (k_4^2 - s)^2 + k_3^2 (6 k_4^2 - 2 s - 8 t) +
      8 (-k_4^2 + s) t + 8 t^2) -
   2 k_2^2 s (-3 k_3^4 +}\\[4pt]
   {\quad(k_4^2 - s) (k_4^2 - s - 4 t) + 2 k_3^2 (k_4^2 + s + 2 t)) -
   2 k_1^2 (k_2^2 ((k_3^2 - k_4^2)^2 + 2 (k_3^2 + k_4^2) s - 3 s^2) +}\\[4pt]
  {\quad    s (k_3^4 - 3 k_4^4 + 2 k_4^2 s + s^2 + 2 k_3^2 (k_4^2 - s - 2 t) +
         4 (k_4^2 + s) t)))\big].
         }
\label{N11value}
\end{array}
\end{equation}
Bear in mind that $s$ and $t$ are the Mandelstam variables $s=(\vec{k}_1+\vec{k}_2)^2$ and $t=(\vec{k}_1+\vec{k}_3)^2$.

Analogously,
\begin{equation}
{\cal W}^{(2)}_s=-\cfrac{L^2}{2 s^2}\,( N_{21}\, \text{J}_{21}+ N_{22}\, \text{J}_{22}+ N_{23}\, \text{J}_{23}+ N_{24}\, \text{J}_{24}),
\label{W2value}
\end{equation}
where
\begin{equation}
\begin{array}{l}
{
N_{21}=\cfrac{1}{16}\,\big[-k_2^4 (k_3^4 - 2 k_3^2 (k_4^2 - 3 s) + (k_4^2 - s)^2) +
   s^2 (-(k_3^2 - k_4^2)^2 + (k_3^2 + k_4^2) s) -}\\[4pt]
  {\quad\quad k_1^4 (k_3^4 + k_4^4 + 6 k_4^2 s + s^2 - 2 k_3^2 (k_4^2 + s)) +
   k_2^2 s (-6 k_3^4 + 2 k_4^4 + s^2 -
   }\\[2pt]
   {\quad\quad k_4^2 (3 s + 8 t) +
      k_3^2 (4 k_4^2 + 5 s + 8 t)) +
   k_1^2 (2 k_2^2 ((k_3 - k_4)^2 + s) ((k_3 + k_4)^2 + s) +}\\[4pt]
{\quad\quad      s (2 k_3^4 - 6 k_4^4 + s^2 + k_3^2 (4 k_4^2 - 3 s - 8 t) +
         k_4^2 (5 s + 8 t)))\big],}\\[4pt]
{
N_{22}= -\cfrac{1}{8}\, (k_1^2 + k_2^2 - s)\, s^2,\quad N_{23}= -\cfrac{1}{8}\, (k_3^2 + k_4^2 - s)\, s^2,\quad N_{24}= -\cfrac{3}{4}\, s^2.
}
\label{N2values}
\end{array}
\end{equation}
The symbols $\text{J}_{2b}$, $b=1,2,3,4$ denote the following integrals:
\begin{equation}
\begin{array}{l}
{
\text{J}_{21}=\int_{0}^{\infty}\! d\omega\;\cfrac{1}{\omega}\;f_2(\omega;k_1,k_2)\,f_{2}(\omega;k_3,k_4),\quad
\text{J}_{22}=\int_{0}^{\infty}\! d\omega\;\cfrac{1}{\omega}\;f_2(\omega;k_1,k_2)\,f_{1}(\omega;k_3,k_4),
}\\[4pt]
{
\text{J}_{23}=\int_{0}^{\infty}\! d\omega\;\cfrac{1}{\omega}\;f_1(\omega;k_1,k_2)\,f_{2}(\omega;k_3,k_4),\quad
\text{J}_{24}=\int_{0}^{\infty}\! d\omega\;\cfrac{1}{\omega}\;f_1(\omega;k_1,k_2)\,f_{1}(\omega;k_3,k_4).
}
\label{Jotados}
\end{array}
\end{equation}
See (\ref{f1}) and (\ref{f2}) for the definition of the functions $f_1$ and $f_2$ in the previous equations. The calculation of the integrals in (\ref{Jotados}) yields
\begin{equation}
\begin{array}{l}
{
\text{J}_{21}=
\cfrac{1}{(k_1 + k_2) (k_3 + k_4) (k_1 + k_2 + k_3 + k_4)^3}\big[k_1^3 (k_3 + k_4) + k_2 (k_2 + k_3 + k_4)}\\[2pt]
   {\phantom{\text{J}_{21}=\,}(k_3^2 + 3 k_3 k_4 + k_4^2 + k_2 (k_3 + k_4))+
k_1 (4 k_2 + k_3 + k_4) (k_3^2 + 3 k_3 k_4 + k_4^2 + }\\[2pt]
{\phantom{\text{J}_{21}=\,}k_2 (k_3 + k_4)) +
   k_1^2 (4 k_2 (k_3 + k_4) + (2 k_3 + k_4) (k_3 + 2 k_4))\big],
   }\\[4pt]
{
\text{J}_{22}=\cfrac{ k_3^2 k_4^2 (k_1^2 + k_2 (k_2 + k_3 + k_4) + k_1 (5 k_2 + k_3 + k_4))}{(k_1 + k_2) (k_1 + k_2 + k_3 + k_4)^4},
}\\[4pt]
{
\text{J}_{23}=\cfrac{ k_1^2 k_2^2 (k_3^2 + k_4 (k_4 + k_1 + k_2) +
   k_3 (5 k_4 + k_1 + k_2))}{(k_3 + k_4) (k_1 + k_2 + k_3 + k_4)^4},
}\\[4pt]
{
\text{J}_{24}=\cfrac{2 k_1^2 k_2^2 k_3^2 k_4^2}{(k_1 + k_2) (k_3 + k_4) (k_1 + k_2 + k_3 + k_4)^3}.
}
\label{Jota2values}
\end{array}
\end{equation}

We shall handle next the computation of ${\cal W}^{(3)}_s$ in (\ref{WGR123def}):
\begin{equation}
{\cal W}^{(3)}_s=-\cfrac{L^2}{2 s^2}\,( N_{31}\, \text{J}_{31}+ N_{32}\, \text{J}_{32}+ N_{33}\, \text{J}_{33}+ N_{34}\, \text{J}_{34}),
\label{W3value}
\end{equation}
with
\begin{equation}
\begin{array}{l}
{
N_{31}=\cfrac{1}{16}\, s\, ((k_1^2 - k_2^2)^2 - (k_1^2 + k_2^2) s) ((k_3^2 -
     k_4^2)^2 - (k_3^2 + k_4^2) s),}\\[4pt]
{
N_{32}=-\cfrac{1}{8}\, s^2 ((k_1^2 - k_2^2)^2 - (k_1^2 + k_2^2) s) ,\; N_{33}=-\cfrac{1}{8}\, s^2 ((k_3^2 - k_4^2)^2 - (k_3^2 + k_4^2) s) ,\; N_{34}=\cfrac{1}{4}\, s^3 .
}
\label{N3values}
\end{array}
\end{equation}
The symbols $\text{J}_{3b}$, $b=1,2,3,4$ stand for the following integrals:
\begin{equation}
\begin{array}{l}
{
\text{J}_{31}=\int_{0}^{\infty}\! d\omega\;\cfrac{1}{\omega^3}\;f_2(\omega;k_1,k_2)\,f_{2}(\omega;k_3,k_4),\quad
\text{J}_{32}=\int_{0}^{\infty}\! d\omega\;\cfrac{1}{\omega^3}\;f_2(\omega;k_1,k_2)\,f_{1}(\omega;k_3,k_4),
}\\[4pt]
{
\text{J}_{33}=\int_{0}^{\infty}\! d\omega\;\cfrac{1}{\omega^3}\;f_1(\omega;k_1,k_2)\,f_{2}(\omega;k_3,k_4),\quad
\text{J}_{34}=\int_{0}^{\infty}\! d\omega\;\cfrac{1}{\omega^3}\;f_1(\omega;k_1,k_2)\,f_{1}(\omega;k_3,k_4).
}
\label{Jotatres}
\end{array}
\end{equation}
See (\ref{f1}) and (\ref{f2}) for the definition of the functions $f_1$ and $f_2$ in the previous equations. After working out the integrals in (\ref{Jotatres}), one gets
\begin{equation}
\begin{array}{l}
{
\text{J}_{31}=
\cfrac{1}{(k_1 + k_2)^3 (k_3 + k_4)^3 (k_1 + k_2 + k_3 +
    k_4)^3}
    \big[k_1^4 (k_3^2 + 3 k_3 k_4 + k_4^2) +
   k_2^2 (k_2 + k_3 +
      k_4)
}\\[2pt]
{
\phantom{\text{J}_{31}=}
      (k_2 (k_3^2 + 3 k_3 k_4 + k_4^2) + (k_3 + k_4) (k_3^2 + 4 k_3 k_4 +
         k_4^2)) +
   3 k_1 k_2 (2 k_2 + k_3 + k_4)
   }\\[2pt]
  {\phantom{\text{J}_{31}=}
       (k_2 (k_3^2 + 3 k_3 k_4 + k_4^2) + (k_3 + k_4) (k_3^2 + 4 k_3 k_4 +
         k_4^2)) +
   k_1^3 (6 k_2 (k_3^2 + 3 k_3 k_4 + k_4^2) +
    }\\[2pt]
 {
 \phantom{\text{J}_{31}=}
   (k_3 + k_4) (2 k_3^2 + 7\, k_3 k_4 +
         2 k_4^2)) +
   k_1^2 (10\, k_2^2 (k_3^2 + 3 k_3 k_4 + k_4^2) +
    }\\[2pt]
   {
   \phantom{\text{J}_{31}=}
   (k_3 + k_4)^2 (k_3^2 +
         4 k_3 k_4 + k_4^2) +
      3 k_2 (k_3 + k_4) (3 k_3^2 + 11 k_3 k_4 + 3 k_4^2))\big],
}\\[4pt]
{
\text{J}_{32}=\cfrac{1}{(k_1 +
   k_2)^3 (k_3 + k_4)^3 (k_1 + k_2 + k_3 + k_4)^3}\,\big[
   (k_3^2 k_4^2 (k_1^4 + 3\, k_1^3 (2\, k_2 + k_3 + k_4) +}\\[2pt]
   {\quad\quad\quad
   k_2^2 (k_2 + k_3 + k_4) (k_2 + 2\,(k_3 + k_4)) +
   3\, k_1 k_2 (2 k_2 + k_3 + k_4) (k_2 + 2\, (k_3 + k_4)) +}\\[2pt]
   {\quad\quad\quad
   k_1^2 (10\, k_2^2 + 15\, k_2 (k_3 + k_4) + 2\, (k_3 + k_4)^2))\big],
}\\[4pt]
{
\text{J}_{33}=\cfrac{1}{(k_3 +
   k_4)^3 (k_1 + k_2)^3 (k_1 + k_2 + k_3 + k_4)^3}\,\big[
   (k_1^2 k_2^2 (k_3^4 + 3\, k_3^3 (2 k_4 + k_1 + k_2) +}\\[4pt]
   {\quad\quad\quad
   k_4^2 (k_4 + k_1 + k_2) (k_4 + 2\, (k_1 + k_2)) +
   3\, k_3 k_4 (2\, k_4 + k_1 + k_2) (k_4 + 2\, (k_1 + k_2)) +}\\[4pt]
   {\quad\quad\quad
   k_3^2 (10\, k_4^2 + 15\, k_4 (k_1 + k_2) + 2\, (k_1 + k_2)^2))\big],
}\\[4pt]
{
\text{J}_{34}=\cfrac{2\, k_1^2 k_2^2 k_3^2 k_4^2 (k_1^2 + 2\, k_1 k_2 + k_2^2 + 3\, k_1 (k_3 + k_4) +
   3\, k_2 (k_3 + k_4) + (k_3 + k_4)^2)}{(k_1 + k_2)^3 (k_3 + k_4)^3 (k_1 + k_2 +
   k_3 + k_4)^3}.
}
\label{Jota3values}
\end{array}
\end{equation}

Now, by substituting (\ref{N11value}) and (\ref{i41}) in (\ref{W1value}), (\ref{N2values}) and (\ref{Jota2values}) in (\ref{W2value}), and
(\ref{N3values}) and (\ref{Jota3values}) in (\ref{W3value}), one obtains ${\cal W}^{(1)}_s$, ${\cal W}^{(2)}_s$ and ${\cal W}^{(3)}_s$, respectively.
Finally, the substitution of the quantities so obtained  in (\ref{WGRtotal}) yields the results quoted in (\ref{WGRND}), (\ref{Dterm}) and  (\ref{Nterm}).

\end{document}